\documentclass[11pt]{article}
\usepackage{comment}
\usepackage[utf8]{inputenc}
\usepackage[margin=1in]{geometry}
\usepackage{subcaption}
\usepackage{graphicx} 
\usepackage{subcaption}

\usepackage{array,multirow}
\usepackage{natbib}
\usepackage{amssymb, amsmath}
\usepackage{amsfonts}
\usepackage{url}

\usepackage{amsthm}

\newtheorem{lemma}{Lemma}
\newtheorem{proposition}{Proposition}

\usepackage{booktabs}
\usepackage[table]{xcolor}
\usepackage{tikz}
\tikzset{
	vertex/.style = {
		circle,
		fill            = black,
		outer sep = 2pt,
		inner sep = 1pt,
	}
}
\usetikzlibrary{chains, positioning, shapes.symbols}

\definecolor{darkred}{RGB}{139,0,0}

\tikzset{
  vertex/.style = {circle, fill=black, outer sep=2pt, inner sep=1pt},
  start/.style  = {signal, fill=#1, minimum height=12mm, draw=white,
                    font=\sffamily\Huge, text=white, inner sep=6pt,
                    signal pointer angle=150, on chain},
  cont/.style   = {start=#1, signal from=west},
  invisible/.style={opacity=0},
  visible on/.style={alt={#1{}{invisible}}},
  highlight on/.style={alt={#1{fill=darkred!80!black}{fill=gray!30!white}}},
  alt/.code args={<#1>#2#3}{%
    \alt<#1>{\pgfkeysalso{#2}}{\pgfkeysalso{#3}}
  }
}
\usepackage{pgfplots}
\usepackage{titling}
\pgfplotsset{compat=1.18}
\usepgfplotslibrary{fillbetween}
\usetikzlibrary{patterns,fillbetween}
 
\pgfplotsset{compat=1.17}

\usepackage{titling}
\usepackage[colorlinks=true,linkcolor = blue, citecolor = blue,  urlcolor = black]{hyperref}

\title{Spiral of Silence: \\ How Neutral Moderation Polarizes Content Creation
\thanks{
\protect\linespread{1}\protect\selectfont 
The authors are listed alphabetically. We are grateful for the feedback received from participants at POMS, IIOC, the Marketing Science Conference, the CESifo Summer Institute on Digital Platforms, the Marketing Theory Seminar, and seminars at the University of Washington and University of Illinois at Urbana-Champaign. We also thank Rafael Jiménez-Durán, David Godes, Michael Keane, Peter Landry, Alexei Makarin, Cameron Martel, Sridhar Moorthy, Marcel Preuss, Mengze Shi, Shubhranshu Singh, Mateusz Stalinski, Michael Zhang, Yanhui Wu, and Pinar Yildirim for helpful conversations. All errors are our own.}
}

\date{\small This Version: November, 2025}

\author{Ying Bao \\ \small{University of Illinois Urbana-Champaign} 
\and Jessie Liu \\ \small{Johns Hopkins University} }

\begin{document}

\maketitle

\begin{abstract}
{\small This paper investigates how content moderation affects content creation in an ideologically diverse online environments. We develop a model in which users act as both creators and consumers, differing in their ideological affiliation and propensity to produce toxic content. Affective polarization, i.e., users' aversion to ideologically opposed content, interacts with moderation in unintended ways. We show that even ideologically neutral moderation that targets only toxicity can suppress non-toxic content creation, particularly from ideological minorities. Our analysis reveals a content-level \textit{externality}: when toxic content is removed, non-toxic posts gain exposure. While creators from the ideological majority group sometimes benefit from this exposure, they do not internalize the negative spillovers, i.e., increased out-group animosity toward minority creators. This can discourage minority creation and polarize the content supply, ultimately leaving minority users in a more ideologically imbalanced environment: a mechanism reminiscent of the ``spiral of silence.'' Thus, our model offers an alternative perspective to a common debate: what appears as bias in moderation needs not reflect bias in rules, but can instead emerge endogenously as self-censorship in equilibrium. We also extend the model to explore how content personalization interacts with moderation policies.}  

\vspace{1em}


\end{abstract}

\clearpage

\setcounter{page}{1}
\pagenumbering{arabic}

\newpage
\section{Introduction}

\begin{quote}
    ``The opinion of only part of the population seemed to be the opinion of all and everybody, and exactly for this reason seemed irresistible to those who were responsible for this deceptive appearance.''\\
    \vspace{1mm}
    \hfill --- Alexis de Tocqueville, \textit{L’Ancien Régime et la Révolution} (1856, p. 259)
\end{quote}

Toxicity is a pervasive challenge on social media platforms, affecting millions of users daily. According to the \cite{Pew2021}, 4 in 10 Americans have faced online harassment, and 71\% support tighter platform rules on toxic content. This growing concern is reflected in emerging regulatory frameworks worldwide, including German Network Enforcement Act (NetzDG) and the European Union’s Digital Services Act (DSA).\footnote{\url{https://digital-strategy.ec.europa.eu/en/library/code-conduct-countering-illegal-hate-speech-online}} 
Existing research on content moderation largely focuses on the demand side: how users respond when speech is restricted. Prior work has shown that content removal can trigger backlash \citep{jhaver2019did}, increase engagement \citep{jimenez2021economics}, or shift user composition \citep{liu2022implications}. While these studies have advanced our understanding of moderation’s impact on content consumption, they often overlook a supply-side question: How the \textit{anticipation} of moderation shapes what content is produced in the first place?

Our study takes this supply-side perspective as its starting point. This focus is increasingly vital in today’s social media landscape. Moderation policies are no longer perceived as occasional or exceptional: they are institutionalized, expected, and often legally mandated \citep{andres2021combating}. As platforms implement preemptive tools such as automated flagging and removal, and as creators adapt to avoid penalties or backlash, anticipatory behavior becomes more relevant than reactionary responses alone. In this regime, it is crucial to understand not only which content gets removed, but also what content is \textit{never created} at all.

A key motivation for our approach comes from a puzzling empirical pattern: opposite moderation policies can sometimes produce similar outcomes. When Reddit banned several toxic communities during ``The Great Ban" in 2020, some former toxic users became more active, but most, especially the less-toxic ones, posted less or stopped altogether \citep{cima2024investigating, cima2024great}. Conversely, when Twitter relaxed its rules in 2022, hate speech nearly doubled \citep{hickey2023auditing}. These two cases illustrate that moderation does more than remove harmful content: it changes the entire environment in which users create content. On social media, users are not merely passive consumers of content, they are also strategic creators who weigh the risks and rewards of content creation. 

Despite growing attention to polarization, most existing research conflate ideological division with toxicity, treating both as a single notion of ``offensiveness.'' This theoretical conflation obscures the distinct incentives and consequences that arise from ideology-based versus toxicity-based preferences. In practice, however, algorithm-based moderation systems mainly targets the latter. For instance, widely used Google's Perspective API are designed to be ideologically neutral, trained on data across the political spectrum and calibrated to detect toxic language rather than viewpoint \citep{rieder2021fabrics}. Yet, content that is non-toxic in language can still provoke severe backlash when it expresses ideological disagreement, depending on \textit{who} is reading. For example, TikTok influencer Leo Skepi, with over 4 million followers, faced strong backlash after stating that brands should not be blamed for not carrying all sizes.\footnote{\url{https://time.com/6965324/leo-skepi-tiktok-clothing-size}}

Disentangling these two types of preferences is therefore conceptually important, not only to understand how users engage with content, but also how they decide \textit{what} to create. Our paper formalizes this distinction along two aspects of user preferences: (i) a \textit{vertical} preference capturing aversion to overtly toxic content, and (ii) a \textit{horizontal} preference capturing aversion to ideologically opposed content. This framework allows us to identify the conditions under which ex ante ``neutral'' moderation, by reweighting anticipated engagement, can produce systematically non-neutral outcomes ex post.  

In this paper, we develop a model in which users act as both consumers and (potential) creators of content, differing along two dimensions: their ideological position and their propensity to produce toxic content.  Content creation depends on both intrinsic motivation and external engagement, whether positive or negative, from ideologically aligned (in-group) and opposing (out-group) audiences. A central feature of our framework is explicitly modeling \textit{affective polarization}: the extent to which users respond emotionally to the ideological identity of content sources, rather than to the content's tone or language itself \citep{iyengar2019origins}. Affective polarization gives rise to systematic \textit{in-group favoritism} and \textit{out-group animosity}: toxic posts often receive some tolerance within ideological groups, whereas even civil content can trigger backlash simply because they come from the ``other side.'' Recent experimental studies also confirm this growing trend: users tend to evaluate otherwise identical content more negatively when it comes from ideological out-group members \citep{wuestenenk2025influence}. Alongside this, our model captures how \textit{ideological imbalance}, i.e., unequal group sizes within the user base, creates asymmetry in who receives more validation versus hostility. Together, these two forces give rise to what we term as \textit{relational externality}: any policy that alters exposure, such as moderation or personalization, reshapes the balance between in-group favoritism and out-group animosity that governs creators’ incentives. In other words, a policy aimed at regulating one dimension of content (e.g., toxicity) can indirectly influence other aspects (e.g., ideology) of content creation by altering the social (relational) environment in which content is rewarded or penalized.

Our analysis offers two key insights. 
First, when both affective polarization and ideological imbalance are high, moderation amplifies in-group favoritism and out-group animosity through increasing the content reach of non-toxic posts from all ideological groups. Facing a larger in-group, this motivates the ideological majority creators to produce more content, whereas the ideological minority are discouraged from creating, as they expect an intensified animosity from a larger out-group. This dynamic reproduces a self-reinforcing ``spiral of silence'' \citep{noelle1974spiral}: what appears as ideological bias in moderation outcomes can emerge \textit{endogenously} in equilibrium, even when moderation rules are designed to be ideologically neutral. This insight also challenges the very notion of ``neutrality'' in policy design. Content-neutral moderation may not be outcome-neutral because it changes the relational environment that determines equilibrium content creation. 

Second, moderation may improve average outcomes but redistributes welfare unevenly across consumers: while all consumers benefit from reduced exposure to toxic content (a universal gain), the composition of what remains skews toward content from the majority group (a polarizing effect). Hence majority readers are exposed to more ideologically aligned content whereas the minority users encounters the opposite. This asymmetry widens welfare inequality across ideological groups. In other words, majority content gain higher reach without internalizing its negative spillover on the minority. These results have direct implications for policy frameworks such as the European Union’s DSA, which emphasizes fairness and transparency in content moderation. Our findings suggest that such frameworks must move beyond static notions of fairness that focus only on what content is removed or demoted. Without accounting for the distinct roles of ideology and toxicity in user behavior, moderation policies may unintentionally reinforce polarization and marginalize civil under-represented groups.

We also extend the baseline model along four dimensions to test the policy relevance and robustness  of our results. The first two extensions examine content personalization and the presence of ideologically neutral users, which are interventions often proposed as immediate remedies to the spiral of silence. We show that personalization protects minority creators from out-group animosity but narrows their reach, and by the same logic can also revive toxicity within ideological groups. Likewise, while a large group of neutral users can diffuse animosity and promote creation, a small one may deepen silence among those very neutral users meant to bridge ideological divides. The final two extensions introduce alternative motivations for toxic users, allowing them to value either toxic consumption or negative engagement. Across both cases, our core insight holds: moderation continues to shape content creation through the same underlying relational externality.

Our results carry important implications for platform governance and the creator economy. Platforms such as Wattpad or YouTube, whose value depends heavily on sustained creator participation, face a fundamental trade-off: how to disentangle toxicity from ideological disagreement in order to mitigate the externalities of the former on creation shaped by the latter. This trade-off becomes especially acute when the user base is ideologically imbalanced and affective polarization is high. Our model also offers a theoretical foundation for recent efforts to design moderation mechanisms that distinguish toxicity from ideological disagreement \citep{birdwatch2021launch}. However, we highlight a critical caveat: unless such mechanisms are incentive-compatible and elicit truthful reporting, flag-based moderation may be abused and institutionalize a new form of silence: not because content is genuinely harmful, but because it is unpopular with dominant groups and thus more likely to be mislabeled as ``toxic.''

Our paper contributes to three strands of literature. First, it advances research on the negative consequences of social media consumption. Prior work has documented that exposure to hate speech increases offline hate crimes \citep{andres2021combating, muller2021fanning, muller2023hashtag}, while curation algorithms on platforms like Facebook and Twitter can amplify trolling, polarization, and echo chambers \citep{cinelli2021echo, levy2021social, bondi2023privacy, pei2024curation,berman2020curation}.  Although most previous studies have recognized the harms of either explicitly harmful content (vertical preference) or ideology-based backlashes (horizontal preference), these forces are often treated in isolation. In practice, they are two aspects manifested in the same content. One exception is \citet{berman2020curation}, which highlights how algorithmic curation affects both the diversity and quality of content consumed. Our model complements this perspective by identifying a structural ``market failure'' that arises not from misaligned incentives between the platform and users, but from polarized content supply driven by asymmetric in-group versus out-group engagement. This externality persists regardless of \textit{who} controls moderation or personalization.

Second, we make a conceptual contribution to the literature on content moderation by emphasizing the dual role of users as both content consumers and strategic content creators. Existing work has shown that moderation policies can reduce audience engagement with hateful content \citep{thomas2023disrupting} and lead to lower hate-content production online as well as offline harm \citep{andres2021combating, jimenez2024effect}.  
Meanwhile, these interventions have been shown, in some cases, to reinforce echo chambers and reduce overall engagement \citep{huang2024politically}. However, these average effects often obscure heterogeneity in user responses. Our study builds on this foundation by endogenizing content creation decisions across different user types. Rather than examining only how users respond to moderated content, we focus on how the \textit{anticipation} of moderation, through its effects on  expected reach and engagement, reshapes the supply of content. This perspective helps reconcile some seemingly divergent empirical findings on how moderation affects content creation: what may appear as null or negative effects on average can, in fact, emerge from offsetting behavioral changes across users with different ideological identities and toxicity.

Third, we formally identify the role of affective polarization in a core marketing context: social media engagement. While affective polarization has been extensively studied in political science \citep{iyengar2019origins, druckman2019we}, its implications for marketing remain underexplored \citep{godes2019politics}. 
As partisan, racial, and religious identities increasingly converge, individuals are more likely to react emotionally to ideologically opposing content \citep{iyengar2019origins}. The rise of partisan media further reinforces these group identities, making individuals more sensitive to perceived in-group and out-group cues, even when their core beliefs remain unchanged \citep{lelkes2017hostile}.
Affective polarization generalizes beyond the simple in-group/out-group dichotomy in political debates and can be viewed as a ``structural property of social networks'' \citep{lerman2024affective}. Our model captures this network feature by linking users' creation and engagement incentives to the relational environment shaped by others’ reactions. This distinction is particularly relevant in digital marketing, where identity signaling and perceived group affiliation are often inseparable from content consumption and creation. For example, affective polarization can intensify backlash to brand activism \citep{homroy2023political} or complicate influencer partnerships when perceived affiliations diverge from audience values \citep{budlight2023}. Our framework offers a tractable approach for marketing scholars to model these features and examine how polarization interacts with platform design, ultimately affecting consumer engagement and brand outcomes.

The rest of the paper is organized as follows. Section \ref{sec:model} introduces the model and equilibrium concept. In Section \ref{sec:analysis}, we assess the impact of moderation. In Section \ref{sec:extention}, we explore a few important extensions of the baseline model. Section \ref{sec: conclusion} concludes with policy and managerial implications.

\section{Model}\label{sec:model}

A social media platform hosts a population of users with a total mass of 1. Each user plays a \textit{potential} dual role as both a content creator and a content consumer — producing content for others to engage with, while also reading and interacting with content created by others. Users differ in two dimensions: their ideology type $i \in \{A, B\}$ and toxicity type $t \in \{T, NT\}$. The ideology type $i$ reflects the view points they lean toward, such as Democrat vs. Republican, or pro-vaccine vs. vaccine-hesitant. The toxicity type reflects their propensity to post toxic ($t = T$) or non-toxic ($t = NT$) content. 
Here, we model toxicity as an exogenous and fixed consumer type, consistent with prior psychology literature that link online trolling to stable personality traits.  For instance, \cite{buckels2014trolls} and \cite{craker2016dark} show that toxic online behavior is strongly associated with enduring dark traits such as sadism, suggesting that the \textit{differential} propensity to engage in toxicity reflects dispositional rather than situational factors.
 We assume that each user can create up to one piece of content that matches with their type.

We introduce a parameter $\delta \in [0, \frac{1}{2}]$ to capture the degree of \textit{ideological imbalance} in the population of platform users. It measures the degree of asymmetry in ideological group size. Specifically, the total shares of users with ideology $A$ and $B$ are given respectively by $(\frac{1}{2} + \delta)$ and $(\frac{1}{2}-\delta)$. In other words, ideology $A$ group is assumed to represent the majority group. This assumption is innocuous, as we remain agnostic about which group is more prone to toxicity — both are treated symmetrically by construction. Let $x \in (0,1)$ denote the overall mass of toxic users in the population. Let $\tau_A \in (0,1)$ and $\tau_B = 1 - \tau_A$ denote the share of toxic users who belong to ideological groups $A$ and $B$, respectively. For example, when $\tau_A = 1$, all toxic users are from group $A$; when $\tau_A = 0$, all toxic users are from group $B$. 
Hereafter in the discussion, for clarity, we always refer to group $A$ as the (ideological) majority group and $B$ as the minority group. 

\begin{table}[h!]
\centering
\begin{tabular}{lcc>{\columncolor[gray]{0.9}}c}
\toprule
 & Toxic ($T$) & Non-toxic ($NT$) & \textbf{Total}\\
\midrule
Ideology $A$ & $x \cdot \tau_A$ & $(\frac{1}{2} + \delta) - x \cdot \tau_A$ & $\frac{1}{2}+\delta$\\
Ideology $B$ & $x \cdot \tau_B$ & $(\frac{1}{2}-\delta) - x \cdot \tau_B$ & $\frac{1}{2}-\delta$ \\
\hline
\rowcolor[gray]{0.9} \textbf{Total} & $x$ & $1-x$ & 1\\
\bottomrule
\end{tabular}
\caption{Mass of User Type by Ideology and Toxicity}
\label{tab:user-mass}
\end{table}

In the subsequent sections, we use the 2-tuple $(i,t)\in\Theta\equiv\left\{ A,B\right\} \times\left\{ T,NT\right\}$ to denote a user's type. We use $\lambda(i,t)$ to denote the population share of a type $(i,t)$ user. Table \ref{tab:user-mass}  summarizes the resulting population mass of each user type.

\subsection{Content Consumption}
A reader of type $(i, t)$ on the platform is exposed to content generated by other users. Their utility from consuming a particular piece of content depends on two key factors: whether the content aligns with the reader’s ideological orientation and whether it contains toxic elements. We represent the utility of a reader $r$ of type $(i, t)$ from consuming content created by a creator $c$ of type $(i’, t’)$ as:

\begin{equation}
U^r(r=it,c=i't')=\underbrace{\alpha \cdot H(i,i')}_{\text{horizontal (ideology-based)}} + \underbrace{(1-\alpha) \cdot V(t')}_{\text{vertical (toxicity-based)}} + \varepsilon_r,
\end{equation}
where 
\begin{align*}
H(i,i') &= \begin{cases}
0 & \text{if }i=i'\\
-1 & \text{if }i\neq i' \end{cases}, \mbox{\ \ \ \quad  }
V(t') =\begin{cases}
0 & \text{if }t'= NT\\
-1 & \text{if }t'= T \end{cases}.
\end{align*}
This utility specification\footnote{The additive structure is a simplifying assumption for analytical tractability and conceptual clarity. It provides a micro-foundation for the probabilistic model of user engagement later described in Table~\ref{tab:engagement_probability}.} captures two dimensions of content evaluation commonly observed on social media platforms. The first, referred to as the \emph{horizontal} dimension ($H(\cdot)$), reflects the tendency for users to disfavor ideologically opposing content, consistent with evidence that users are more receptive to viewpoints that match their own \citep{kozyreva2023resolving}. The second, the \emph{vertical} dimension, reflects a general aversion to harmful or toxic content, independent of ideological alignment. Note that, in our baseline model, the utility specification implies that a user’s own toxicity type $t\in\{NT,T\}$ does not affect their utility from consuming content, i.e., $U^r(it,i't')=U^r(i,i't')$.  This reflects the assumption that users evaluate others’ content based on its ideology and tone, rather than their \textit{own} posting behavior. In other words, we treat content toxicity as a vertical attribute in the baseline model: users who engage in toxic creation also experience disutility from consuming toxic content, particularly when it originates from the ideological out-group \citep{rabbani2025explaining}. In Section \ref{sec:TTH}, we relax this assumption of toxicity as a purely vertical attribute and allow toxic users to derive utility from toxicity.

The parameter $\alpha \in [0,1]$ calibrates the relative weight users place on ideological alignment versus content toxicity. One can also interpret $\alpha$ as the degree of aversion to the opposing ideology. When $\alpha$ is close to 1, users prioritize ideological alignment over toxicity concerns. This formulation resonates with the political science literature on \emph{affective polarization}: the increasing animosity between the parties, even in the absence of substantive ideological divergence. As discussed by \citet{iyengar2019origins}, affective polarization stems from partisanship functioning as a social identity, where individuals categorize others into a favored in-group and a disfavored out-group, independent of policy-specific disagreement (p. 130).
Finally, $\varepsilon_r \sim U[-1,1]$ denotes users' idiosyncratic utility shock from consuming content.

Upon consuming content created by others, a reader can decide whether to engage with the content by liking or disliking the content. The like and dislike decisions do not necessarily mean the like button or dislike button. Instead, we use them to represent all positive and negative engagement with the posts, including reactive emojis and comments directed towards the posts.  Research suggests that social media users are generally more inclined to express positive feedback, such as “likes,” rather than negative feedback, such as “dislikes.” This tendency is influenced by platform design, such as the prominent display of like buttons, and psychological factors, including the drive for social validation \citep{stsiampkouskaya2023like}. Thus, we assume that a reader will like a post if the utility from consuming it is greater than 0, i.e. $U^r\geq 0$, whereas they will dislike a post only if the utility is below a threshold, i.e., $U^r\leq -\gamma$. Here, $\gamma\in(0,1)$ can be considered as the relative cost of negative engagement: the higher $\gamma$ is, the less likely a reader will dislike the content. Figure \ref{fig:engagement} below illustrates readers' engagement pattern induced by their utility $U^r(\cdot)$.

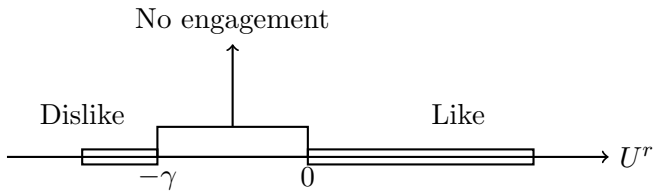
\begin{figure}[h!]
		\centering
		\begin{tikzpicture}
			\draw[thick,->] (-4,0) -- (4,0) node[right] {$U^r$}; 
			\draw[thick,->] (-1,0.4) -- (-1,1.5) node[above] {No engagement}; 
			
			\node[below] at (0,0) {$0$};
			\node[below] at (-2,0) {$-\gamma$};
			
			\draw[thick] (-3,0.1) -- (-2,0.1) -- (-2,-0.1) -- (-3,-0.1) -- cycle; 
			\draw[thick] (0,0.1) -- (3,0.1) -- (3,-0.1) -- (0,-0.1) -- cycle; 
			
			\node[above] at (-3,0.3) {Dislike};
			\node[above] at (2,0.3) {Like};
			
			\draw[thick] (-2,0) rectangle (0,0.4);
			
		\end{tikzpicture}
    \caption{User Engagement Based on $U^r(\cdot)$}
    \label{fig:engagement}
\end{figure}

For simplicity, we normalize $\gamma$ to $\frac{1}{2}$ in the main model.\footnote{Note that this normalization does not qualitatively change our results as long as it is exogenously given between 0 and 1.} As discussed above, the readers' toxicity type does not affect their consumption utility, i.e., $U^r(it,i't')=U^r(i,i't'), \forall t\in\{NT,T\}$. 
Thus, we use $P_l(r=i,c=i't')$ and $P_{dl}(r=i,c=i't')$ to denote respectively the probability of positive and negative engagement of a reader $r$ consuming a post created by a creator $c$. Given the utility specification, Table \ref{tab:engagement_probability} summarizes these probabilities in different cases.   

\begin{table}[h!]
\begin{minipage}{0.45\textwidth}
\centering
\begin{tabular}{@{}cccccc@{}}
    \toprule
    \multirow{2}{*}{$P_l(r,c)$} & \multicolumn{4}{c}{\textbf{Creator}} \\ \cmidrule(lr){2-6}
     && $A,NT$ &$ A,T$ & $B,NT$ & $B,T$ \\ \midrule
    \multirow{2}{*}{\textbf{Reader}} &$ A$ & $\frac{1}{2}$ & $\frac{\alpha}{2}$ & $\frac{1-\alpha}{2}$ & $0$ \\
    & $B$ & $\frac{1-\alpha}{2}$ & $0$ & $\frac{1}{2}$ & $\frac{\alpha}{2}$ \\ 
    \bottomrule
\end{tabular}
\vspace{0.5em}

(a) Positive Engagement
\end{minipage}
\hspace{0.04\textwidth}
\begin{minipage}{0.45\textwidth}
\centering
\begin{tabular}{@{}cccccc@{}}
    \toprule
    \multirow{2}{*}{$P_{dl}(r,c)$} & \multicolumn{4}{c}{\textbf{Creator}} \\ \cmidrule(lr){2-6}
     && $A,NT$ & $A,T$ & $B,NT$ & $B,T$ \\ \midrule
    \multirow{2}{*}{\textbf{Reader}} & $A$ & $\frac{1}{4}$ & $\frac{3-2\alpha}{4}$ & $\frac{1+2\alpha}{4}$ & $\frac{3}{4}$ \\
    & $B$ & $\frac{1+2\alpha}{4}$ & $\frac{3}{4}$ & $\frac{1}{4}$ & $\frac{3-2\alpha}{4}$ \\
    \bottomrule
\end{tabular}
\vspace{0.5em}

(b) Negative Engagement
\end{minipage}
\vspace{1em}
\caption{Positive (a) and Negative (b) Engagement Probabilities by Reader and Creator Type}

\label{tab:engagement_probability}
\end{table}

Taken together, these probabilities suggest that users are more likely to positively engage with (i.e., ``like'') non-toxic content created by users with the same ideology. Meanwhile, users are more likely to negatively engage with (i.e., ``dislike'') toxic content from creators  with the opposite ideology. These behavioral patterns also align with the empirical findings documented in \cite{kozyreva2023resolving}.

\subsection{Content Creation and Moderation}

In addition to reading others’ posts, users can also create content. Prior work suggests that content creation is motivated by both intrinsic utility and the desire for external recognition or engagement \citep{toubia2013intrinsic}. We formalize the utility of content creation for a user (creator) $c$ of type $(i,t)$ as follows:
\begin{equation}
    U^{c}(i,t)=\underbrace{S\left(t\right)}_{\text{survival rate}}\cdot\underbrace{R(V)}_{\text{content reach}}\cdot\underbrace{(NE^{in}(i,t)+NE^{out}(i,t))}_{\text{in- and out-group engagement}}\underbrace{+\varepsilon_{c}}_{\text{intrinsic utility}},
\end{equation}
where $\varepsilon_c \sim U[-1,1]$ captures idiosyncratic or intrinsic motivations unrelated to external reactions. The first three terms jointly capture the expected utility from external engagement, shaped by the following three factors:
\begin{enumerate}
\item 
	Content Survival ($S(t)$): Content must survive moderation to be able to reach audiences. We model moderation as acting on posts rather than on reactions. This matches empirical practice: hate speech and toxicity are measured and reported primarily at the level of top-level posts, as in \cite{hickey2023auditing}'s analysis of toxic tweet spikes and \cite{muller2021fanning,muller2023effects}'s measurement of hate tweets. Moreover, \cite{beknazar2022toxic} show that widely used moderation tools, such as Perspective-based filters, operate mainly through post-level adjustments, whereas replies receive far less exposure and are not the primary target of automated moderation.\footnote{Allowing moderation on reactions would not materially alter our qualitative results. Moderating negative reactions effectively removes \textit{some} negative engagement that involves uncivil behavior. Incorporating reaction-level moderation would therefore resemble a reduction in $|\omega|$, i.e., the discouraging effect from negative engagement, and the core polarizing mechanism we highlight would continue to hold under the relevant parameter ranges.} 
    Let $\beta \in [0,1]$ denote the intensity of the platform’s moderation policy, representing the share of toxic posts removed before exposure. We define survival probability as:
\[S(t) = 1 - \beta \cdot \mathbf{1}[t = T],\]
i.e., only toxic posts are subject to removal. This formulation allows for partial enforcement, reflecting regulatory limits driven by free speech concerns or technical feasibility \citep{dave2020social,carlson2020report}. A regime with full enforcement ($\beta = 1$) removes all toxic posts, whereas $\beta = 0$ implies no moderation. A post fails to go live with probability $1 - S(t)$.

\item 
	Content Reach ($R(V)$): Once live, content competes for user attention.   
    We assume that each surviving post attains a uniform reach across the platform's readership. Section~\ref{sec: personalization} relaxes this assumption by introducing content personalization. The platform’s exposure mechanism is modeled through a linear ``reach factor'' ($R(\cdot)$) that decreases with the total content supply ($V$): 
    \begin{equation}
R(V) = 1 - V; \quad V=\sum_{i,t\in\{A,B\}\times\{T,NT\}} \lambda(i,t) \cdot S(t) \cdot P_c(i,t),
\label{eq:visibility}
\end{equation}
    where $\lambda(i,t)$ denotes the population share of type $(i,t)$ users (as shown in Table \ref{tab:user-mass}), and $P_c(i,t)$ is the expected share of content created by such users.  This formulation captures how total content supply reduces the expected exposure of any \textit{single} post \citep{iyer2016competing}. It reflects a well-documented empirical regularity: from the creator's perspective, an increase in others’ content reduces the likelihood that their own post is seen. For instance, 
    
    Facebook’s ranking pipeline trims “thousands of candidate posts” to just a few hundred, necessarily reducing exposure as content supply grows \citep{meta2021}. The linear specification keeps creators’ utilities affine in $P_c$, enabling closed-form equilibrium solutions. It offers a tractable way to capture this congestion property that higher aggregate activity lowers per-post reach.
   Classical congestion models \citep[e.g.,][]{acemoglu2007competition,johari2010investment} adopt smooth, monotone mappings with the same logic.  To verify that our qualitative results are not driven by this linear approximation, In Appendix, we re-solve the model numerically using three other bounded alternatives, e.g. inverse, exponential, and logistic reach functions, and confirm that all yield qualitatively identical equilibrium patterns.

\item 

	User Engagement ($(NE^{in}(i,t)+NE^{out}(i,t)$): If a post is seen, it may generate engagement from both ideological in-group readers ($NE^{in}(i,t)$) and out-group readers ($NE^{out}(i,t)$). These are calculated as expected responses from all reader types:
\begin{align}
 NE^{in}(i,t) &= \underbrace{\lambda_i}_{\text{in-group share}}\cdot [\underbrace{P_{l}\!\left(r=i,c=it\right)}_{\text{prob. of like}} + \omega \underbrace{P_{dl}\!\left(r=i,c=it\right)}_{\text{prob. of dislike}}], \\
 NE^{out}(i,t) &= \underbrace{\lambda_{-i}}_{\text{out-group share}}\cdot \left[P_{l}\!\left(r=-i,c=it\right) + \omega P_{dl}\!\left(r=-i,c=it\right)\right],
\end{align}

where $P_l$ and $P_{dl}$ denote, from the perspective of a creator $c$ of type $(i,t)$ , the probabilities that their content receives a like or dislike, respectively, from the in-group ($r=i$) or out-group ($r=-i$) readers, whose respective share is denoted by $\lambda_i$ and $\lambda_{-i}$.\footnote{Here, the $\lambda_i$ and $\lambda_{-i}$ can be computed based on Table \ref{tab:user-mass}. } Together with the engagement probabilities from Table \ref{tab:engagement_probability}, our specification implies the ``hallmarks'' of affective polarization, characterized by an emotional divide of \textit{in-group favoritism} and \textit{out-group animosity} \citep{lerman2024affective}. That is, creators, even toxic ones, tend to get more positive reactions from ideologically aligned readers (in-group favoritism) and more negative reactions from ideologically opposing readers (out-group animosity). The parameter $\omega \in [-1, 0)$ captures the discouraging effect of negative engagement on content creation. Equivalently, $|\omega|$ represents the creator's degree of aversion to negative engagement. This is consistent with empirical findings that negative feedback tends to suppress creator activity \citep{berger2012makes, RedditDownvote2025}. We explore the possibility that some users may derive positive utility from negative engagement (i.e., $\omega > 0$) in Section~\ref{sec:toxic_creator_dislike}. 
\end{enumerate}
Finally, a user of type $(i,t)$ will choose to create content if and only if:
$U^c(i,t) > 0$.

\subsection{Equilibrium Concept}

We adopt the concept of rational expectation equilibrium \citep{grossman1980impossibility, moorthy1985using}, in which the expected share of each user type that creates content equals the probability of creation implied by their utility. Formally, the following condition must hold for all user types:
\begin{align} \label{eq:pc0_pc1}
P_c(i,t) = \Pr[U^c(i,t) > 0], \quad \forall i \in \{A,B\}, t \in \{T,NT\}.
\end{align}
In equilibrium, all creators share a common belief about $P_c(i,t)$, which enters the content reach term $R(\cdot)$ and must be internally consistent with the realized outcome. As long as $P_c(\cdot)\in[0,1]$, we have $R(\cdot)\in[0,1]$, ensuring equilibrium existence via Brouwer’s fixed-point theorem. 

In the next section, we discuss how moderation shapes content creation of each types and their respective welfare.

\section{Analysis} \label{sec:analysis}

We begin our analysis with two benchmark cases: (i) no affective polarization ($\alpha = 0$) and (ii) affective polarization present ($\alpha>0$) with a balanced ideological composition ($\delta = 0$). 
We then relax these constraints to explore the full parameter space, considering environments characterized by both affective polarization among users ($\alpha > 0$) and ideological imbalance between groups ($\delta > 0$).
By comparing equilibrium outcomes in case (ii) to case (i), we isolate the effect of vertical, toxicity-based preferences from that of horizontal, ideology-based preferences. This comparison  demonstrates the role of affective polarization in shaping creators' incentives under content moderation. Comparing the full model to case (ii) allows us to examine how the effects of both horizontal and vertical preferences are further amplified or mitigated when one ideological group dominates the platform. Throughout our analysis, we maintain the assumption that the total share of toxic users is not too large ($x \leq \frac{2}{3}$). This restriction simplifies our analysis and helps us focus on equilibrium outcomes that are most relevant in practice, where moderation unambiguously reduces the equilibrium exposure of toxic content. In Appendix, we show that there exists a unique equilibrium and provide the detailed equilibrium characterization. 

Proofs for all subsequent result and propositions are also included in the Appendix.

\begin{lemma}[\textbf{Affective Polarization without Group Imbalance}]\label{result:alpha0}
When ideological groups are balanced ($\delta=0$), the effect of moderation depends entirely on affective polarization ($\alpha$). For all $i\in\{A,B\}$, $\beta\in[0,1]$, $\tau_A\in(0,1)$:

\begin{itemize}
\item[(a)] Neutral Environment: 
 When $\alpha=0$, moderation unambiguously increases non-toxic content creation while reduces survived toxic content across both ideological groups:
\[
\frac{\partial P_c(i,NT)}{\partial \beta}>0,
\quad
\frac{\partial [S(T)\cdot P_c(i,T)]}{\partial \beta}<0,
\]
\item [(b)] Polarized Environment: 
When $\alpha>0$ and creators are highly averse to negative feedback ($\omega\le -\tfrac{1}{2}$), stronger affective polarization can sometimes make moderation counterproductive:  
\[\frac{\partial P_c(i,NT)}{\partial \beta}
\begin{cases}
<0 & \text{if } \alpha>\alpha_1(\omega)\equiv \tfrac{2+\omega}{1-\omega},\\[3pt]
\ge 0 & \text{otherwise},
\end{cases}
\qquad
\qquad \frac{\partial [S(T)\cdot P_c(i,T)]}{\partial \beta}<0.
\]
\end{itemize}
\end{lemma}

In the first benchmark case without affective polarization ($\alpha=0$), readers evaluate content solely by its toxicity rather than ideology. From a creator’s perspective, expected engagement therefore depends only on whether their post is toxic or non-toxic, making the effect of moderation symmetric across ideological groups. Removing toxic posts increases the content reach of surviving material because fewer posts compete for reader attention. This broader reach encourages non-toxic creators from both groups to produce more. Meanwhile, the total volume of toxic content that survives moderation declines monotonically.

When affective polarization emerges ($\alpha>0$), it changes creation incentives. Readers begin to display in-group favoritism and out-group animosity: they react more favorably to ideologically aligned content and more negatively to opposing views. As a result, moderation’s impact depends critically on the degree of affective polarization. With moderate $\alpha$, creation incentives remain roughly balanced across groups. However, when affective polarization is high and creators are highly averse to negative feedback, moderation can become counterproductive: even as it reduces surviving toxic posts, it can discourage the creation of non-toxic users, thereby shrinking the pool of civil content that moderation aims to promote. Conditional on exposure, users' aversion to negative engagement ($\omega$) amplifies the harm of animosity relative to the benefit of favoritism, whether from in-group or out-group members; As $\alpha$ increases, out-group tolerance weakens while their animosity intensifies; once $\alpha$ exceeds a critical threshold $\alpha_1(\omega)$, the expected disutility from out-group animosity outweighs the utility from in-group favoritism, which is amplified by increased content reach to both ideological groups due to moderation. Consistent with this mechanism, \citet{thomas2022s} report that roughly 22\% of creators self-censor content (about themselves or their beliefs) to avoid negative reactions, illustrating how anticipated backlash can deter creation.

Next we turn to the case where both ideological imbalance ($\delta>0$) and affective polarization ($\alpha>0$) are present. It allows us to identify how unequal group sizes and preference for ideological alignment jointly shape the effects of content moderation. We can show that the total volume of toxic content that survives moderation continues to decline monotonically, i.e., $\frac{\partial [S(T)P_c^*(i,T)]}{\partial \beta}<0$, consistent with Lemma~\ref{result:alpha0}. 
In the proposition below, we therefore focus on how moderation changes the incentives of \textit{non-toxic} creators, who play a central role in maintaining long-term content quality and user retention on the platform.

\begin{proposition}[\textbf{Spiral of Silence Equilibrium}]\label{prop:pc0_delta}
  For $\beta\in[0,1]$, $\tau_A\in(0,1)$, moderating toxic content affects the equilibrium content creation among non-toxic creators from ideological groups A and B differently. Specifically, three distinct regions emerge:
\begin{itemize}
    \item[(a)]\textbf{Universal Suppression}: If $\alpha > \bar{\alpha}$, moderation reduces non-toxic content creation from both groups, i.e., $\frac{\partial P_{c}^{*}(A,NT)}{\partial \beta}<0, \frac{\partial P_{c}^{*}(B,NT)}{\partial \beta}<0.$
    
    \item[(b)]\textbf{Universal Empowerment}: If $\alpha < \underline{\alpha}$, moderation increases non-toxic content creation from both groups, i.e.,
    $\frac{\partial P_{c}^{*}(A,NT)}{\partial\beta}>0, \quad \frac{\partial P_{c}^{*}(B,NT)}{\partial\beta}>0$.

    \item[(c)]\textbf{Polarized Creation}: If $\underline{\alpha} \leq \alpha \leq \bar{\alpha}$, moderation polarizes creation: majority group creates more whereas minority group creates less, i.e.,
    $\frac{\partial P_{c}^{*}(A,NT)}{\partial\beta}\geq 0, \frac{\partial P_{c}^{*}(B,NT)}{\partial \beta} \leq 0$.
  \end{itemize}
Specifically, $\underline{\alpha} = \frac{\omega +2}{(1+2 \delta ) (1-\omega )}$ and $\bar{\alpha} = \frac{\omega +2}{(1-2 \delta ) (1-\omega )}$. 
\end{proposition}

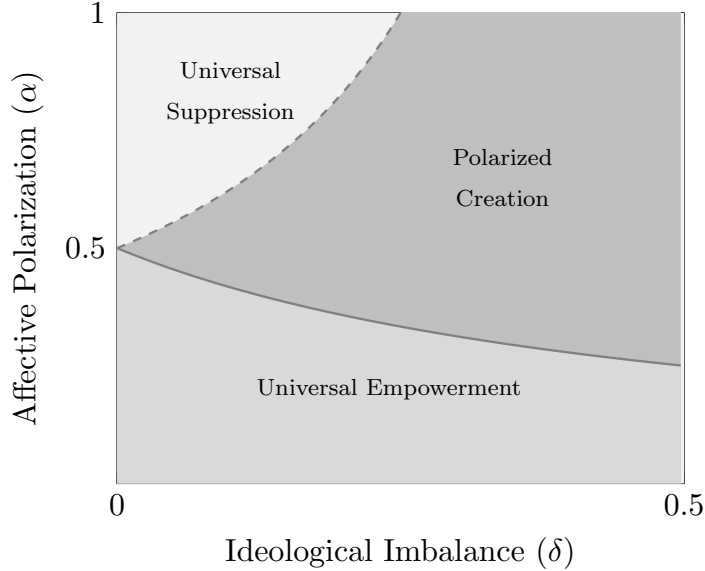
\begin{figure}[h!]
  \centering
  \begin{tikzpicture}[scale=1.1]
    \begin{axis}[
      ylabel={Affective Polarization ($\alpha$)},
      xlabel={Ideological Imbalance ($\delta$)},
      ymin=0, ymax=1,
      xmin=0, xmax=0.5,
       xtick={0, 0.5, 1},       
  ytick={0.5, 1}, 
      axis lines=box,
      unbounded coords=jump,
    ]
      \addplot[gray!10, fill=gray!10, draw=none]
        coordinates {(0,0) (0.5,0) (0.5,1) (0,1)} \closedcycle;

      \addplot[gray!50, fill=gray!50, draw=none, domain=0.0001:0.497, samples=200]
        {1/(2*(1-2*x))} \closedcycle;

      \addplot[gray!30, fill=gray!30, draw=none, domain=0.0001:0.497, samples=200]
        {1/(2*(1+2*x))} \closedcycle;

\addplot[gray, dashed, thick, domain=0.0001:0.497, samples=200]
  {1/(2*(1 - 2*x))};

\addplot[gray, thick, domain=0.0001:0.497, samples=200]
  {1/(2*(1 + 2*x))};
 
  \node at (axis cs:0.24, 0.2) [
    font=\scriptsize , 
    color=black, 
    text width=3.3cm, 
    align=center
  ] {Universal Empowerment};
  
  \node at (axis cs:0.34, 0.65) [
    font=\scriptsize , 
    color=black, 
    text width=2.2cm, 
    align=center
  ] {Polarized Creation};
  
  \node at (axis cs:0.1, 0.83) [
    font=\scriptsize, 
    color=black, 
    text width=2.2cm, 
    align=center
  ] {Universal Suppression};

\end{axis}
  \end{tikzpicture}
  \caption{Effect of Moderation on Non‐toxic Content Creation ($\omega=-1$). \\
  {\footnotesize Note: $\alpha\in [0,1]$  reflects the degree of affective polarization, i.e., the weight users place on ideology relative to toxicity (higher $\alpha$ means stronger aversion to opposing ideology). $\delta\in[0,0.5]$ captures ideological imbalance across users, ranging from 0 (equal group sizes) to 0.5 (one group dominates the entire market). The dashed line represents the upper threshold, $\overline{\alpha}$, and the solid line represents the lower threshold, $\underline{\alpha}$. }}
  \label{fig:pc0_delta}
\end{figure}


Proposition \ref{prop:pc0_delta} shows that when one ideological group dominates the population, moderation no longer affects creators uniformly across ideological lines,  for any relative share of toxic users across groups ($\forall\tau_A\in(0,1)$).  Instead, its impact depends critically on both the degree of affective polarization ($\alpha$) and the extent of ideological imbalance ($\delta$). As illustrated in Figure \ref{fig:pc0_delta}, three distinct equilibrium regions emerge.
When affective polarization is weak ($\alpha < \underline{\alpha}$), we are in the ``universal empowerment'' region: moderation enhances the reach of non-toxic content, and the benefits outweigh the costs for non-toxic creators across both groups. This logic echoes that of Lemma \ref{result:alpha0}(a).
At the other extreme, when affective polarization is strong ($\alpha > \bar{\alpha}$), we enter the ``universal suppression'' region: moderation triggers a negative feedback loop driven by intensified out-group animosity, ultimately reducing content creation from both sides. This rationale aligns with Lemma \ref{result:alpha0}(b).
Between these two extremes lies the ``polarized creation'' region, which arises under moderate affective polarization ($\underline{\alpha}\leq \alpha \leq \overline{\alpha}$). Here, moderation amplifies existing ideological imbalances: creators from the ideological majority group are encouraged to produce more, while those from the minority group withdraw. The intuition is as follows: moderation removes toxic content, increasing the content reach of non-toxic posts from both groups. However, the resulting engagement dynamics differ sharply: ideological majority creators anticipate more in-group favoritism, while minority creators expect more out-group animosity, which can be strong enough to outweigh the support from their own smaller in-group base.

The final result from Proposition \ref{prop:pc0_delta} echoes the classic ``spiral of silence'' effect in social psychology, first described by \citet{noelle1974spiral} as a process through which majority voices dominate public discourse while minority voices fall silent, ``increasingly establishing the [majority] opinion as the prevailing one'' (p.~44). Traditionally, this spiral is interpreted as a psychological reaction to social pressure. Our model reveals a different, equilibrium-based logic: content moderation can \textit{induce} rather than enforce silence in equilibrium. By changing creator incentives through anticipated content reach, even ideologically neutral moderation could discourage creation from ideological minorities and generate the appearance of bias without any explicit censorship.

Our findings also shed new lights on a long-standing debate: while some commentators attribute higher takedown rates to ideological ``censorship,'' some empirical studies often point instead to “toxicity asymmetry,” where certain ideological groups host more toxic content \citep{haimson2021disproportionate}. We offer an alternative explanation. Even when the number of toxic users is identical across groups, stricter moderation can \textit{differentially} suppress or amplify non-toxic content, driven entirely by supply-side responses to expected engagement. Thus, the observed asymmetry, where one side appears to ``lose'' more content, may not reflect inherent toxicity. Rather, it may emerge endogenously from how moderation policies interact with ideological composition and creator incentives. In this sense, what critics call ``censorship'' and what empirical studies term ``toxicity asymmetry'' could be seen as two sides of the same equilibrium process: one that transforms externally neutral moderation into self-censorship.

Building on Proposition \ref{prop:pc0_delta}, which shows how moderation differentially affects content creation across ideological groups, we now turn to its welfare implications. Rather than analyzing each group in isolation, we focus on the welfare inequality between two groups, i.e., $\mathbb{E}[U^r(A)-U^r(B)]$. This metric clarifies how moderation policies shape relative outcomes between ideological groups. 

For clarity and simplicity, in the following proposition and extension sections, we assume an equal mass of toxic users across ideological groups, i.e., $\tau_A = \tau_B = \frac{1}{2}$. This is because when either $\tau_A$ or $\tau_B$ dominates, the result on welfare inequality is intuitive as the dominating group will bear most of the impact of content moderation. Moreover, this constraint helps disentangle the effects of toxicity and ideology by removing any built-in correlation between the two in user composition. Our setup allows us to examine how ideological imbalance ($\delta$) and affective polarization ($\alpha$) shape moderation outcomes, independent of any ideological differences in toxicity.\footnote{Our result is robust to the case where $\tau_A \neq \tau_B$, as long as the gap $\tau_A - \tau_B$ is not too large. A formal proof is available upon request.} In doing so, we also shed new light on the above debate over whether ideological differences in toxicity actually matter. \label{note: tauA=tauB}

\begin{proposition}[\textbf{Welfare Redistribution}] \label{prop:welfare}
Stricter moderation (larger $\beta$) enlarges readers' welfare gap across two ideology groups. That is, for any $\alpha>0, \delta>0$, we have:
\[\frac{\partial\,\mathbb{E}[U^r(A)-U^r(B)]}{\partial\beta}>0,\forall \beta\in[0,1].\]
\end{proposition}

Proposition \ref{prop:welfare} shows that moderating toxic content redistributes reader welfare across ideological groups. The intuition naturally follows Proposition \ref{prop:pc0_delta}: although both groups benefit from reduced exposure to toxic content (a universal gain), the composition of what remains tilts toward content from the majority group (a polarizing effect). As a result, majority readers are exposed to more ideologically aligned content whereas the minority users face the opposite. This asymmetry drives an increasing slope in welfare inequality.

Once again, our findings from Propositions \ref{prop:pc0_delta} and \ref{prop:welfare} underscore a fundamental challenge for platform governance: moderating offensive content without undermining ideological diversity. Viewed through a Coasean lens \citep{coase1960problem}, moderating toxicity generates externalities that disproportionately burden non-toxic ideological minorities. When toxic content is removed, non-toxic posts gain greater exposure. Majority-group creators benefit from this shift but do not internalize the negative spillovers, namely, increased out-group animosity directed at minority creators. This reduces minority creation and polarizes content supply, leaving minority readers in a more ideologically imbalanced environment. To internalize this externality, platforms may benefit from designing mechanisms, such as flagging systems that distinguish between offensive content and ideological disagreement, that help disentangle toxicity aversion from affective polarization. In fact, some platforms have begun allowing users to categorize content as ``toxic language,'' ``misinformation,'' or ``offensive ideology.'' For example, Twitter's  Birdwatch (now Community Notes) program enables users to label and contextualize misleading or offensive tweets \citep{birdwatch2021launch}. 

However, such systems remain vulnerable to strategic misreporting. To address this distortion, platforms must implement incentive-compatible mechanisms that promote truthful reporting. Without such design, flag-based moderation risks institutionalizing a spiral of silence, not because content is \textit{actually} harmful, but because it is unpopular with dominant groups. In such cases, the negative externalities of toxicity become increasingly difficult to disentangle from ideological disagreement. While a full mechanism design is beyond the scope of this paper, our model suggests that platforms should aim to elicit private information about the true \textit{intent} behind negative engagement, rather than simply suppressing tools like the ``thumbs down'' or dislike button, which may redirect backlash to the comment section \citep{kim2024youtube}. 

Finally, our results challenge the common assumption that content-neutral moderation is ideologically neutral in its effects. Even when moderation targets only toxicity, the resulting shifts in content reach and engagement can systematically disadvantage ideological minorities. This concern is further complicated in community-moderated platforms like Reddit, where moderation is decentralized and subreddit norms vary widely. For instance, \cite{rajadesingan2021political} show that politically oriented subreddits often apply rules asymmetrically, with content from ideological out-groups more likely to be flagged or removed. Additionally, users may strategically report ideologically opposing content to suppress dissent, a phenomenon often observed during polarized events such as elections or protests. Therefore, platform design must go beyond neutrality in policy and audit moderation decisions for group-level fairness, not just accuracy.

In the next section, we develop several important extensions to our baseline model. They serve to further illustrate the limitations of moderation policies that disregard \textit{relational} externalities.

\section{Extension} \label{sec:extention}
In this section, we explore four extensions of the baseline model.  The first introduces content personalization that tailors exposure to ideologically aligned audiences. The second adds a group of ideologically neutral users who are non-partisan and care only about content toxicity. We examine these two extensions first because they are often viewed as quick remedies to the distortions implied by the baseline model: reducing cross-group exposure or neutralizing some users' ideology. Yet, as we show in this section, their effects are more nuanced than these intuitive prescriptions suggest.

By contrast, the last two extensions focus on the alternative motivations for toxic users, first as readers who derive utility from seeing toxic content directed at ideological opponents, and then as creators who value negative engagement. These variations introduce richer behavioral complexity while demonstrating the robustness of our main results. In both cases, the qualitative patterns from the baseline model remain intact, suggesting that our core mechanism does not depend on the specific assumption about toxic users' motivations.

\subsection{Content Personalization}
\label{sec: personalization}
Our baseline model assumes that a creator’s post is displayed uniformly across the platform such that the expected reach for a single post is identical across two partisan groups. In practice, however, social media platforms frequently personalize exposure based on users' ideological affinity \citep{gonzalez2023asymmetric,eg2023scoping}. To capture this feature, we extend the model by allowing \textit{differential} content reach and engagement between ideological in-group and out-group audiences. This extension clarifies how the platform’s personalization design interacts with moderation outcomes. 

We begin from the reader’s perspective. A reader of ideology $i$ is exposed to both in-group ($i'=i$) and out-group ($i'=-i$) content of toxicity type $t\in\{NT,T\}$. Now the respective exposure probabilities are given by:
\begin{align}
P_r^{in}(i,t)&=\frac{ \lambda(i,t)S(t)P_{c}(i,t)}{\sum_{t} \lambda(i,t)S(t)P_{c}(i,t) + (1-\phi)\lambda(-i,t)S(t)P_{c}(-i,t)};\\
P_r^{out}(i,t)&=\frac{ (1-\phi)\lambda(-i,t)S(t)P_{c}(-i,t)}{\sum_{t} \lambda(i,t)S(t)P_{c}(i,t) + (1-\phi)\lambda(-i,t)S(t)P_{c}(-i,t)},
\end{align}
where $\phi\in[0,1]$ calibrates how much personalization filters out cross-group exposure. When $\phi=0$, content reach of both groups is fully symmetric as in the baseline model; higher values of $\phi$ indicate stronger personalization toward the creator's in-group audience via reduced cross-group exposure.
Accordingly, the creator’s expected utility is modified as follows:
\[
U^{c}(i,t)
= S(t)\cdot
\Big[ R^{in}\cdot NE^{in}(i,t) + R^{out}\cdot\, NE^{out}(i,t)\Big],
\]
where the reach of a single post to in-group and out-group readers is modified as follows:\footnote {Note that per-post reach $R^{in}, R^{out}$ and realized exposure shares $P_r^{in}, P_r^{out}$ operate at different aggregation levels and represent distinct perspectives on content exposure, though they remain internally consistent. Specifically, $R(\cdot)$ reflects the ex-ante (expected) reach of a \textit{single} surviving post from the creator’s perspective, capturing competition for limited attention, whereas $Pr(\cdot)$ denotes the ex-post share of \textit{total} exposures attributed to content of a given type from the reader’s perspective.} 
\begin{align}
R^{in} =1-V, 
\quad R^{out}=(1-\phi)(1-V); \quad
V  =\sum_{(i,t)\in \Theta} \lambda(i,t) \cdot S(t) \cdot P_c(i,t).
\end{align}
The modified terms, $R^{in}$ and $R^{out}$, refer to the content reach to in-group and out-group readers, respectively, where cross-ideological reach is restricted by the degree of personalization $\phi$.  
All other aspects of the model remain unchanged. 

In the baseline model, we show that minority creators may withdraw when moderation intensifies, as their content faces stronger out-group animosity. This raises a natural question: could content personalization, by exposing users to less ideologically opposing content, help mitigate this spiral of silence? To examine this possibility, we focus on how personalization changes creation incentives for minority creators. The result is summarized in the following proposition.

\begin{proposition} (\textbf{content personalization and moderation})\label{prop:person}
\noindent
\begin{enumerate}
\item[(a)] When $\phi <\underline{\phi}=\frac{4\delta\left(1-\omega\right)+4\omega+2}{3\omega(1-2\delta)}$, our main results regarding non-toxic user's content creation ($\frac{\partial P_c^*(A,NT)}{\partial  \beta}$ and $\frac{\partial P_c^*(B,NT)}{\partial  \beta}$) remain qualitatively the same. 
\item[(b)] Under a given moderation policy ($\beta>0$), personalization may decrease or increase content creation by non-toxic minority creators. In particular,  if $\alpha<\underline{\alpha}(\beta, \phi,x, \delta)$, $\frac{\partial P_c^*(B,NT)}{\partial \phi} <0$, otherwise, $\frac{\partial P_c^*(B,NT)}{\partial \phi} >0$. Meanwhile, under insufficiently strict moderation ($\beta <\underline{\beta}(\alpha,\delta,\phi,x)$), the surviving minority-toxic content increases with personalization, i.e., $\frac{\partial[(1-\beta)P_c^*(B,T)]}{\partial \phi}>0$.
\end{enumerate}
\end{proposition}

Proposition~\ref{prop:person} shows that when personalization remains moderate ($\phi < \underline{\phi}$), our main qualitative results continue to hold. Meanwhile, it also suggests that personalization and moderation are neither perfect substitutes nor complete remedies for the externalities identified in the baseline model. Moderation removes toxic content directly, while personalization redirects exposure toward ideologically aligned readers. For minority creators, these two forces operate in tension: personalization provides only conditional relief for minority non-toxic users, while weakening the disciplining effect of moderation on minority-toxic ones.

Specifically, a high degree of personalization introduces a distinct trade-off for non-toxic creators. On one hand, personalization shields minority creators from the out-group animosity that moderation amplifies in the baseline model. By concentrating exposure within their own ideological group, minority creators anticipate less negative engagement and relatively more positive engagement. This protection can offset the discouragement caused by moderation, leading some non-toxic minority creators to maintain or even increase their creation.
On the other hand, personalization also narrows their potential reach: posts are now shown primarily to in-group readers. This reduced audience size weakens the incentive to create, particularly when affective polarization is low and in-group favoritism carries limited emotional payoff. The balance between these two effects determines the slope of minority non-toxic creation with respect to moderation. When affective polarization is mild ($\alpha<\underline{\alpha}$), the loss of reach dominates, so personalization discourages minority non-toxic creation ($\frac{\partial P_c^*(B,NT)}{\partial \phi}<0$). When affective polarization is high, the effect of protective engagement prevails, and personalization can instead encourage minority creation ($\frac{\partial P_c^*(B,NT)}{\partial \phi}>0$). 

Although such personalized exposure may appear beneficial, it still carries the well-documented concern of ``echo chambers,'' where algorithmic curation confines users within ideologically homogeneous environments \citep{cinelli2021echo, huang2024politically, gonzalez2023asymmetric}. Beyond this established concern, however, our model identifies an additional class of risk: when the moderation is insufficiently strict ($\beta<\underline{\beta}$), greater personalization also increases the survival of minority-toxic content ($\frac{\partial[(1-\beta)P_c^*(B,T)]}{\partial \phi}>0$). 
By promoting exposure through sympathetic in-group readers, toxic minority creators are motivated to create more as they anticipate less backlash, enabling more of their posts to persist despite moderation.

\begin{figure}[h!]
    \centering
    \begin{subfigure}[t]{0.48\textwidth}
        \centering
        \includegraphics[scale=0.45,trim=1cm 6.5cm 1.2cm 6.5cm,clip]{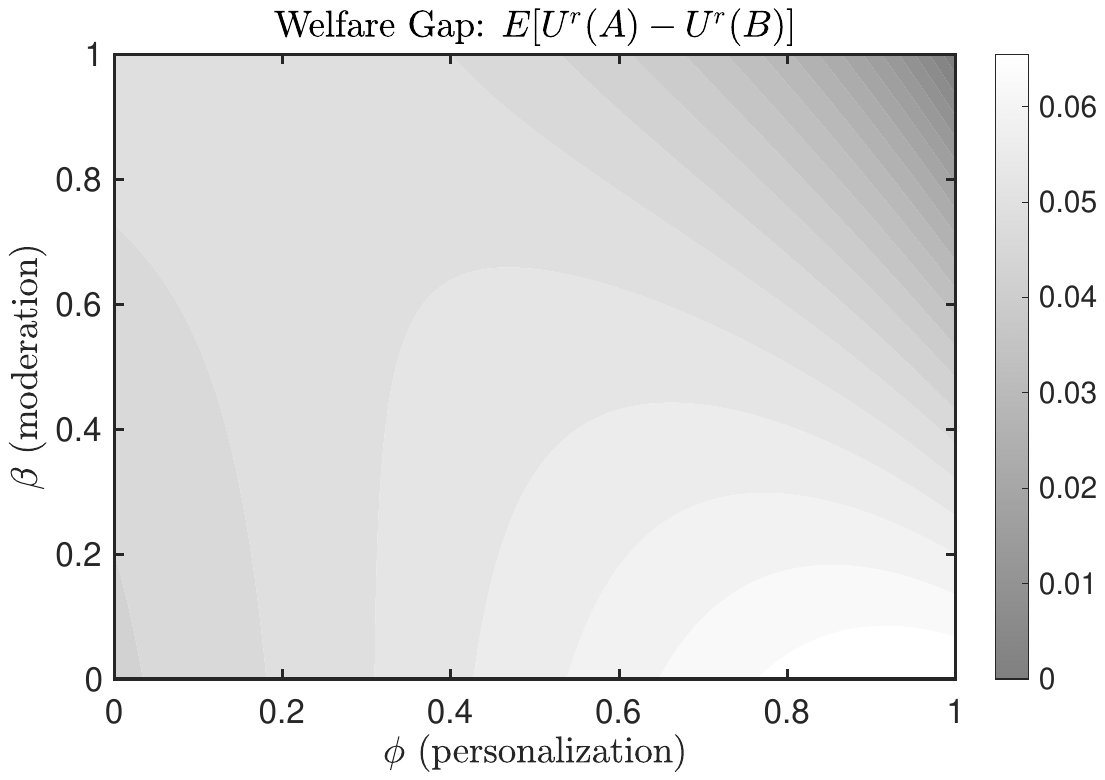}
        \caption{$\alpha = 0.2, \delta =0.1$ }
        \label{fig:welfare-gap-person-small}
    \end{subfigure}
    \hfill
    \begin{subfigure}[t]{0.48\textwidth}
        \centering
        \includegraphics[scale=0.45,trim=1cm 6.5cm 1.2cm 6.5cm,clip]{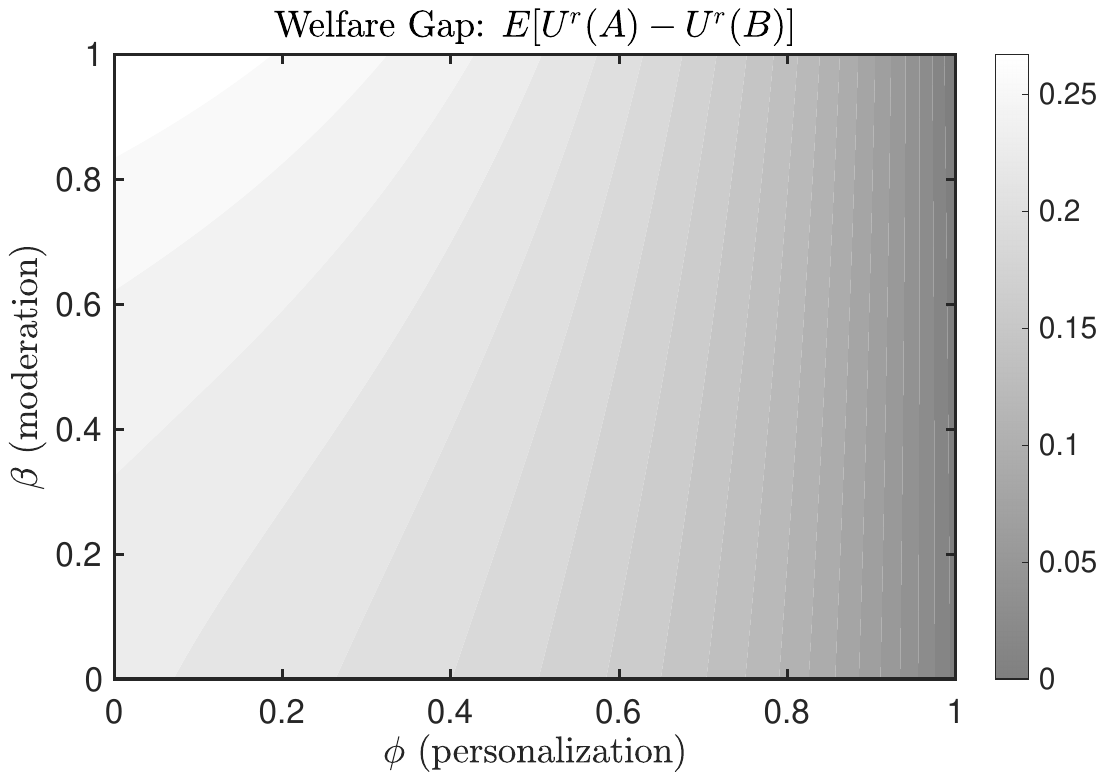}
        \caption{$\alpha = 0.9,\delta = 0.1$}
        \label{fig:welfare-gap-person_large}
    \end{subfigure}
    \caption{$\mathbb{E}[U^r(A)-U^r(B)]$ over degree of personalization ($\phi$) and moderation intensity ($\beta$)} \label{fig:person}
\end{figure}

To further illustrate how moderation and personalization jointly shape reader welfare, we conduct a numerical analysis under low ($\alpha=0.2$) and high ($\alpha=0.9$) affective polarization. The results, summarized in Figure \ref{fig:person}, reveal that moderation and personalization are neither perfect substitutes nor perfect complements. As shown in Figure \ref{fig:welfare-gap-person-small}, when both moderation and affective polarization are low, personalization can exacerbate welfare inequality, likely due to reduced cross-group reach. In contrast, Figure \ref{fig:welfare-gap-person_large} presents a different pattern: under high affective polarization, personalization helps narrow the welfare gap created by moderation, primarily by mitigating cross-group animosity when it matters most, though this effect diminishes at high levels of moderation intensity.

Taken together, our analysis shows that moderation and personalization interact in non-linear ways. When affective polarization is strong, moderate personalization can complement strict moderation by protecting minority creators from out-group animosity. But when moderation is weak, the same personalization revives toxic creation among the minority. In less polarized environments, personalization not only offers in-group comfort to toxic creators but weakening cross-group validation for non-toxic ones.  In other words, even as content personalization protects the minority creators from out-group animosity, it can simultaneously induce a higher level of toxic content and sometimes even exacerbate the ``spiral of silence'' discussed earlier. This creates a subtler form of polarization beyond the standard echo-chamber explanations. These patterns highlight the importance of evaluating policy fairness at different levels of affective polarization ($\alpha$) and jointly over both moderation and personalization, rather than treating each dimension in isolation.

From a practical standpoint, depending on their user base and mission, platforms face distinct optimal combinations of $(\beta,\phi)$. Highly polarized platforms (e.g., X/Twitter, Truth Social) may benefit from moderate personalization that limits cross-group hostility while maintaining strict moderation to minimize in-group toxicity. More heterogeneous or knowledge-oriented communities (e.g., Reddit, Wikipedia) should resist strong personalization to preserve content diversity and reduce risks of echo-chamber.

\medskip{}

\subsection{Ideological Neutral Users}\label{sec:neutral}

In the baseline model, all users are partisan ($i\in\{A,B\}$). In practice, some users on the platform may not have strong ideology preference. According to the survey by \cite{pew2024partisanship}, around 35\% of the registered voters say they are independent. This extension incorporates that possibility by introducing an ideologically neutral group of users ($i = N$). Their utility function as a reader $r$ seeing a post $c$ is simply given by 
\[U^r(r=N,c=i't')=V(t)+\varepsilon_r. \]
Here, we assume that the neutral readers only care about the toxicity aspect of content. In terms of content creation, for simplicity, we rule out the possibility that neutral users post toxic content. Accordingly, the user type space is defined as $\Theta=\{(N,NT)\}\cup(\{A,B\}\times \{T,NT\})$. Let $\lambda_N\in(0,1)$ denote the mass of neutral users. Accordingly, the mass of $A$-group and $B$-group are given by $(1-\lambda_N)(\frac{1}{2}+\delta)$ and $(1-\lambda_N)(\frac{1}{2}-\delta)$, respectively. Given the addition of neutral (non-toxic) content, the utility function of partisan readers $i\in\{A,B\}$ is now modified as follows: 
\[U^r(r=i,c=i't')=\alpha \cdot H(i,i')+(1-\alpha)\cdot V(t)+\varepsilon_r,\]
where $H(i,i')$ is given by:
\[
H(i,i’)=\begin{cases}
0,& \text{if } i=i',\\[2pt]
-\frac{1}{2},& \text{if } i'=N,\\[2pt]
-1,& \text{otherwise.}
\end{cases}
\]

The horizontal dimension $H(i,i')$ takes three values depending on which group the content is from. We assume that reader prefer content from the same ideology group to neutral group to the opposing group. 
This specification retains the feature of in-group favoritism and out-group animosity from the baseline model, while adding that partisan readers exhibit (symmetrically) moderate tolerance toward ideologically neutral content. All other aspects of the model remain unchanged. The result is summarized in the following proposition.

\begin{figure}[ht]
  \centering
\begin{subfigure}[t]{0.3\textwidth}
  \centering
  \begin{tikzpicture}[scale=0.7]
    \begin{axis}[
      ylabel={Affective Polarization ($\alpha$)},
      xlabel={Ideological Imbalance ($\delta$)},
      ymin=0, ymax=1,
      xmin=0, xmax=0.5,
       xtick={0, 0.5, 1},       
  ytick={0.5, 1}, 
      axis lines=box,
       unbounded coords=jump,
    ]
      \addplot[gray!10, fill=gray!10, draw=none]
        coordinates {(0,0) (0.5,0) (0.5,1) (0,1)} \closedcycle;

      \addplot[gray!50, fill=gray!50, draw=none, domain=0.0001:0.497, samples=200]
        {5/(9*(1-2*x))} \closedcycle;

      \addplot[gray!30, fill=gray!30, draw=none, domain=0.0001:0.497, samples=200]
        {5/(9*(1+2*x))} \closedcycle;

\addplot[gray, dashed, thick, domain=0.0001:0.497, samples=200]
  {5/(9*(1 - 2*x))};

\addplot[gray, thick, domain=0.0001:0.497, samples=200]
  {5/(9*(1 + 2*x))};

 
  \node at (axis cs:0.24, 0.2) [
    font=\scriptsize , 
    color=black, 
    text width=3.3cm, 
    align=center
  ] {Universal Empowerment};
  
  \node at (axis cs:0.3, 0.65) [
    font=\scriptsize , 
    color=black, 
    text width=2.2cm, 
    align=center
  ] {Polarized Creation};
  
  \node at (axis cs:0.08, 0.85) [
    font=\scriptsize, 
    color=black, 
    text width=2.2cm, 
    align=center
  ] {Universal Suppression};

\end{axis}
  \end{tikzpicture}
  \caption{ $\lambda_N=0.1$}
  \label{fig:pc0_delta_neutral_small}
\end{subfigure}
   \quad \quad
\begin{subfigure}[t]{0.3\textwidth}
  \centering
  \begin{tikzpicture}[scale=0.7]
    \begin{axis}[
      xlabel={Ideological Imbalance ($\delta$)},
      ymin=0, ymax=1,
      xmin=0, xmax=0.5,
       xtick={0, 0.5, 1},       
  ytick={ 1}, 
      axis lines=box,
       unbounded coords=jump,
    ]
      \addplot[gray!50, fill=gray!50, draw=none]
        coordinates {(0,0) (0.5,0) (0.5,1) (0,1)} \closedcycle;

      \addplot[gray!30, fill=gray!30, draw=none, domain=0.001:0.497, samples=200]
        {1/(1+2*x))} \closedcycle;

\addplot[gray, thick, domain=0.0001:0.497, samples=200]
  {1/(1 +2*x)};

 
  \node at (axis cs:0.25, 0.4) [
    font=\scriptsize , 
    color=black, 
    text width=3.3cm, 
    align=center
  ] {Universal Empowerment};
  
  \node at (axis cs:0.32, 0.79) [
    font=\scriptsize , 
    color=black, 
    text width=2.2cm, 
    align=center
  ] {Polarized Creation};
  

\end{axis}
  \end{tikzpicture}
  \caption{ $\lambda_N=0.5$}
  \label{fig:pc0_delta_neutral}
\end{subfigure}
\begin{subfigure}[t]{0.3\textwidth}
  \centering
  \begin{tikzpicture}[scale=0.7]
    \begin{axis}[
      xlabel={Ideological Imbalance ($\delta$)},
      ymin=0, ymax=1,
      xmin=0, xmax=0.5,
       xtick={0, 0.5, 1},       
  ytick={1}, 
      axis lines=box,
       unbounded coords=jump,
    ]
      \addplot[gray!30, fill=gray!30, draw=none]
        coordinates {(0,0) (0.5,0) (0.5,1) (0,1)} \closedcycle;



 
  \node at (axis cs:0.245, 0.49) [
    font=\scriptsize , 
    color=black, 
    text width=3.3cm, 
    align=center
  ] {Universal Empowerment};
  
  

\end{axis}
  \end{tikzpicture}
  \caption{ $\lambda_N=
  0.9$}
  \label{fig:pc0_delta_neutral}
\end{subfigure}
  \caption{ Effect of Moderation in the Presence of Neutral Users ($\omega=-1$; $\lambda_N\in\{0.1,0.5,0.9\}$)}
  \label{fig:pc0_delta_neutral}
\end{figure}
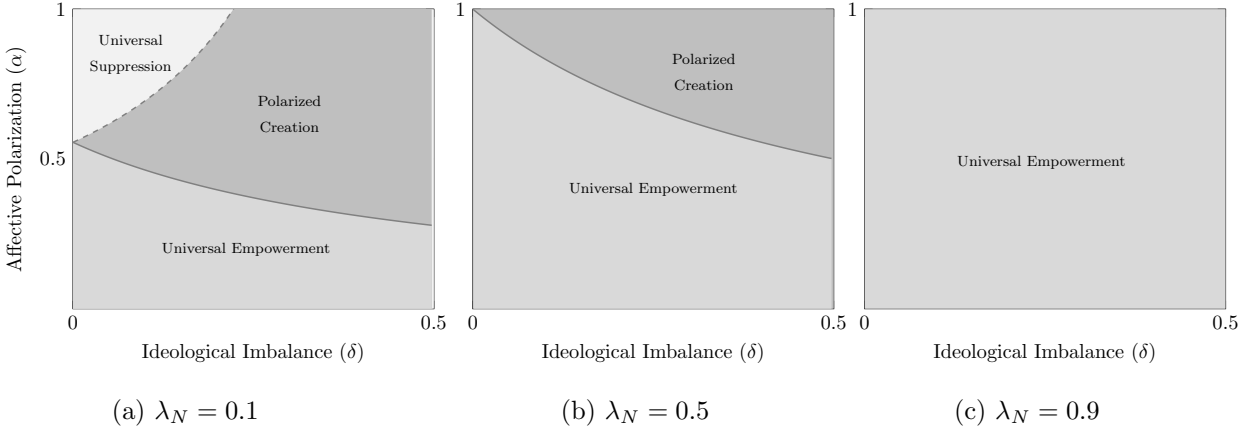

\renewcommand{\labelitemii}{$\bullet$}
\begin{proposition} (\textbf{Ideological Neutral Users})\label{prop:neutral}
  
 \begin{itemize}
     \item[(a)] Moderation increases neutral content if and only if the share of neutral users is large enough, i.e., if $\lambda_N < \lambda_0\equiv2-\frac{3}{1-\omega }$, then $\frac{\partial P_c^*(N,NT)}{\partial \beta}<0$; otherwise, $\frac{\partial P_c^*(N,NT)}{\partial \beta}\geq 0$.
     \item[(b)] Moderation induces the following three equilibrium regions at varying shares of neutral users: 
     \begin{itemize}
        \item  Universal Suppression: when  $\lambda_N<\underline{\lambda_N}$, moderation suppresses the creation of both partisan groups, i.e., $\frac{\partial P_c^*(A,NT)}{\partial \beta}<0, \quad \frac{\partial P_c^*(B,NT)}{\partial \beta}<0$.
        \item Polarized Creation: when $\underline{\lambda_N}\leq \lambda_N \leq\overline{\lambda_N}$, moderation polarizes the creation of partisan groups, i.e.,   $\frac{\partial P_c^*(A,NT)}{\partial \beta}\geq 0, \quad \frac{\partial P_c^*(B,NT)}{\partial \beta}\leq 0$.
        \item Universal Empowerment: when $\lambda_N> \overline{\lambda_N}$,   moderation empowers the creation of both partisan groups, i.e., $\frac{\partial P_c^*(A,NT)}{\partial \beta}>0, \quad \frac{\partial P_c^*(B,NT)}{\partial \beta}>0$.
     \end{itemize}
\end{itemize}
Specifically, $\overline{\lambda_N}\equiv 1-\frac{\omega +2}{\alpha (1-\omega)(1+2\delta)}$ and $\underline{\lambda_N} \equiv 1-\frac{\omega +2}{\alpha (1-\omega)(1-2\delta)}$.
\end{proposition}

Proposition \ref{prop:neutral} shows how the presence of ideologically neutral users affects the impact of moderation on partisan users' content creation incentives. Neutral readers systematically reward non-toxic content, diluting out-group animosity and generating no ideology-based backlash. Proposition \ref{prop:neutral}(a) shows that when such users make up only a small share of the platform, they may become even less active as moderation intensifies. The logic parallels the spiral of silence: with a limited audience base, neutral creators effectively become the new minority, where appreciation from their small in-group is outweighed by partisan indifference or mild hostility. As a result, they are more likely to remain silent despite a declining level of toxicity on the platform. Only once their share exceeds a critical threshold do neutral creators become more active under stricter moderation. In that regime, neutral users not only soften the tension between the two partisan groups but also crowd out their reach. Consequently, the incentives of non-toxic partisan creators depend on the trade-off between expected reach and engagement. 

Proposition \ref{prop:neutral}(b) further shows that when the neutral-user share is very small ($\lambda<\underline{\lambda_N}$), outcomes resemble those in the baseline model. As their share rises to moderate levels ($\underline{\lambda_N}\leq \lambda\leq \overline{\lambda_N}$), their role as an ``animosity absorber'' outweighs the crowd-out effect on majority creators, eliminating the region of universal suppression but still generating a spiral-of-silence equilibrium under high affective polarization. Finally, when neutral users dominate the platform ($\lambda>\underline{\lambda_N}$), cross-group animosity among partisans largely dissipates, and the outcome converges to that in Lemma \ref{result:alpha0}(a). Figure \ref{fig:pc0_delta_neutral} illustrates how the two partisan groups respond to moderation under different shares of neutral users.

From a policy perspective, our findings in this extension suggest that encouraging ideological neutrality among only a few users can deepen their isolation, as they lose audience without changing the broader climate of hostility. Only when neutrality becomes common enough to reshape the overall climate does moderation begin to promote content creation rather than silence.

\subsection{Targeted Toxicity Homophily}
\label{sec:TTH}
In our baseline model, toxicity is treated strictly as a vertical trait, i.e., it is uniformly disliked by everyone. Whereas in practice, some toxic creators may derive satisfaction from toxicity directed at their ideological opponents.  Empirical studies find that ideologically aligned users react more favorably to toxic attacks \textit{on} the out-group \citep[e.g.,][]{yu2024partisanship,lerman2024affective}. To accommodate this interpretation while maintaining the distinction between ideology and toxicity, we extend the reader–creator utility to allow what we refer to as \emph{targeted toxicity homophily}: toxic readers value toxicity when it is targeted toward the opposing group, i.e., when it is aligned with their own ideology.  Formally, the reader $r=(i,t)$'s utility from a creator $c=(i',t')$’s post is now given by:
\[
U^{r}(r=it,c=i't')=\alpha \cdot H(i,i')+(1-\alpha)\cdot \mathcal{T}(it,i't')+\varepsilon_r,
\]
where
\[
H(i,i')=
\begin{cases}
0,& i=i',\\[-2pt]
-1,& i\neq i',
\end{cases}
\qquad
\mathcal{T}(it,i't')=\begin{cases}
0 & t'=NT\\
-1, & t'=T,t=NT\\
\dfrac{\kappa\alpha}{1-\alpha}\cdot{\mathbf{1}[i= i']-1}, & t'=T,t=T
\end{cases},
\]
and $\kappa\in(0,1)$ measures the strength of this targeted homophily. When $\kappa=0$, the model reverts to the baseline in which toxicity uniformly reduces all reader utility; for $\kappa>0$,   toxic readers obtain additional utility from toxic content produced by same-ideology creators, i.e., $\mathcal{T}(iT,iT)=\frac{\kappa \alpha}{1-\alpha}-1$. In contrast, when they encounter toxic content from the opposing group, their utility remain the same as in the baseline model, i.e. $\mathcal{T}(iT,-iT)=-1$. In other words, toxic readers continue to dislike toxicity directed toward their own side but gain satisfaction when toxicity aligns with their in-group ideology. For non-toxic readers, they continue to dislike toxic content regardless of the ideology associated with the creator, consistent with the baseline model.
This extension captures the identity-affirming function of hostility frequently observed in online discourse, where ``punching the other side'' garners approval within the in-group. In this sense, $\mathcal{T}(it,i't')$ reflects the relational nature of toxicity itself within polarized environment. 

All other aspects of the model remain unchanged. The result is summarized in the following proposition. 

\begin{proposition}\label{prop:pc0_delta_homopily}
 Compared to the baseline model, when toxic users value toxic contents from the same group, 
 \begin{itemize}
     \item  the main results regarding non-toxic user's content creation ($\frac{\partial P_c^*(A,NT)}{\partial  \beta}$ and $\frac{\partial P_c^*(B,NT)}{\partial  \beta}$) and welfare inequality $(\frac{\partial \mathbb{E}[U^r(A)-U^r(B)]}{\partial \beta})$ remain qualitatively unchanged, $\forall \beta \in [0,1]$. 
     \item  the (surviving) toxic content increases with the degree of targeted toxicity homophily. That is,  $\frac{\partial [(1-\beta)P_c^*(i,T)]}{\partial \kappa}>0$, $\forall i\in\{A,B\}$, $\kappa \in(0,1)$. 
 \end{itemize}

\end{proposition}

Proposition \ref{prop:pc0_delta_homopily} shows that under targeted toxicity homophily, our main qualitative results remain unchanged: the three equilibrium regions from the baseline model persist. This is because targeted homophily directly shapes how toxic readers engage with toxic content, thereby influencing the incentives of toxic creators. In contrast, it affects \textit{non-toxic} creators only indirectly through its impact on the equilibrium feedback loop. Despite this indirect channel, the expected reach of non-toxic posts still increases with moderation. Thus, the core intuition of the baseline model continues to hold.

The key difference from the baseline model lies in the behavior of toxic rather than non-toxic creators. The homophily term raises toxic creators' expected positive engagement (and dampens expected negative engagement) toward same-side toxic content. Consequently, toxic creators are further incentivized to produce toxic posts. 

\subsection{Toxic Creators Valuing Negative Engagement}
\label{sec:toxic_creator_dislike}
 
In the main model, we assume negative engagement discourages content creation by setting $\omega < 0$. This assumption aligns with many platforms’ efforts to reduce the salience of negative feedback, such as YouTube’s 2021 decision to make dislike counts private. This reflects concerns that visible disapproval deters participation. However, some studies suggest that some users, particularly toxic ones, may be motivated by negative responses. For instance, \cite{buckels2014trolls} find that certain toxic individuals derive pleasure from eliciting anger or outrage. 

To capture this heterogeneity in content creation incentives, we extend our model to allow negative engagement to encourage toxic users while continuing to discourage non-toxic users. Let $\omega(t)$ denote the weight placed on negative engagement by a type-$t$ creator. For illustrative purposes,  we set $\omega(NT) = \omega$ and $\omega(T) = -\omega$, with $\omega \in [-1, 0)$. Recall that the penalty (or reward) from negative engagement, captured by $\omega$, enters the in-group and out-group engagement terms as follows:
\begin{align*}
 NE^{in}(i,t) &= \lambda_{i}\cdot [P_{l}\!\left(r=i,c=it\right)+ \omega(t) \cdot P_{dl}\!\left(r=i,c=it\right)], \\
 NE^{out}(i,t) &= \lambda_{-i}\cdot \left[P_{l}\!\left(r=-i,c=it\right) + \omega(t)\cdot P_{dl}\!\left(r=-i,c=it\right)\right].
\end{align*}
All other aspects of the model remain unchanged. The result is summarized in the following proposition. 

\begin{proposition}\label{prop:pc0_delta_dislike}
 When toxic users value negative engagement, the main results regarding non-toxic user's content creation $( \frac{\partial P_c(A,NT)}{\partial \beta}$ and $\frac{\partial P_c(B,NT)}{\partial \beta})$ and welfare inequality $(\frac{\partial \mathbb{E}[(U^r(A)-U^r(B)]}{\partial \beta})$ remain qualitatively unchanged $\forall \beta \in [0,1]$. In addition, the equilibrium level of  (surviving) toxic content is higher than the main model. 
\end{proposition}

This extension confirms the robustness of our main findings. The intuition is as follows. When toxic users value  negative engagement, their incentives for content creation are directly strengthened. Their impact on non-toxic creators' content creation is only through the change in content supply and corresponding changes in expected reach. We show that 
moderation consistently raises their expected reach and, conditional on reach, their expected engagement remains the same.  Consequently, the strategic environment faced by non-toxic users remains largely unchanged. 

As a result, key outcomes, such as their content creation and welfare inequality, continue to move in the same direction as in the baseline model. As for toxic users, by contrast, valuing negative engagement strengthens their motivation to produce toxic content, leading to a higher level of toxic content in equilibrium.

\section{Conclusion}
\label{sec: conclusion}

This paper offers a strategic perspective on moderating toxicity in online platforms, showing that content moderation is not only about what gets removed, but also shapes what gets created. Importantly, we show that even when moderation targets toxicity alone, their effects may not be ideologically neutral: stricter enforcement can encourage
ideological majority creation at the expense of silencing minority. These asymmetries arise not from explicit bias but from structural externalities in content governance.  
As a result, the same policy can silence, stimulate, or polarize creation depending on the structure of affective divide within its user base.

These findings carry both managerial and policy implications. For platforms that rely on user-generated content, such as Wattpad and YouTube, sustaining active creation requires more than removing harmful content; it demands designing incentive-compatible systems that separate toxicity from ideological disagreement. To complement such mechanisms, platforms may also explore ways to reduce affective polarization across the board. For instance, by elevating norm-setting users with low toxicity and high credibility across ideological lines, or offering prompts that encourage users to frame disagreement constructively. From a regulatory standpoint, our results underscore the need to consider the structural and behavioral roots of polarization when evaluating fairness in content governance. Future work could extend our framework by modeling the design of truthful flagging mechanisms that reduce such negative externalities. 

To highlight the core mechanism, our model deliberately abstracts from two additional forces: endogenous user exit and advertising incentives.  Under certain conditions, incorporating them would likely reinforce rather than overturn our results. Allowing readers to leave the platform when their utility falls below a threshold would disproportionately drive minority attrition, shrinking their audience base and amplifying ideological imbalance, thereby accelerating the spiral toward ``polarized creation.'' Likewise, we omit advertisers to focus on consumer behavioral responses to moderation, but the insights extend naturally to a profit-maximizing platform: when moderation is guided by engagement metrics, it can still magnify inequality in creation and welfare across ideological groups, even under rules designed to be neutral. Future research can formally test these conjectures in settings where both user retention and revenue incentives are endogenous.

Our study also offers a few testable empirical predictions. For instance, we predict that identical moderation policies may produce asymmetric effects on content creation depending on a platform’s ideological composition and the degree of affective polarization. Holding content quality constant, minority-group creators are likely to experience more negative engagement and sharper declines in creation following stricter moderation. We hope future research will examine these predictions in field or experimental settings to guide evidence-based content governance.

\bigskip
\bibliographystyle{apalike}
\bibliography{lit}

@article{rabbani2025explaining,
  title={Explaining the victim-offender overlap of cyberbullying using low self-control and parental bonds},
  author={Rabbani, Md Golam and Pusch, Natasha},
  journal={Crime \& Delinquency},
  volume={71},
  number={10},
  pages={3219--3243},
  year={2025},
  publisher={SAGE Publications Sage CA: Los Angeles, CA}
}

@article{jimenez2024effect,
  title={The effect of content moderation on online and offline hate: Evidence from Germany's NetzDG},
  author={Jim{\'e}nez Dur{\'a}n, Rafael and M{\"u}ller, Karsten and Schwarz, Carlo},
  journal={Available at SSRN 4230296},
  year={2024}
}

@article{stsiampkouskaya2023like,
  title={To like or not to like? An experimental study on relational closeness, social grooming, reciprocity, and emotions in social media liking},
  author={Stsiampkouskaya, Kseniya and Joinson, Adam and Piwek, Lukasz},
  journal={Journal of Computer-Mediated Communication},
  volume={28},
  number={2},
  pages={zmac036},
  year={2023},
  publisher={Oxford University Press}
}

@article{beknazar2022toxic,
  title={Toxic content and user engagement on social media: Evidence from a field experiment},
  author={Beknazar-Yuzbashev, George and Jim{\'e}nez Dur{\'a}n, Rafael and McCrosky, Jesse and Stalinski, Mateusz},
  journal={Available at SSRN 4307346},
  year={2025}
}

@techreport{pei2024curation,
  title={Do Curation Algorithms Amplify the Effect of Trolls on Users?},
  author={Pei, Amy and Mayzlin, Dina},
institution = {Working Paper},
year={2024}
}

@article{jimenez2021economics,
  title={The Economics of Content Moderation: Evidence from Hate Speech on Twitter},
  author={Jim{\'e}nez Dur{\'a}n, Rafael},
  journal={Available at SSRN 4044098},
  year={2021}
}

@article{bondi2023privacy,
  title={Privacy and polarization: An inference-based framework},
  author={Bondi, Tommaso and Rafieian, Omid and Yao, Yunfei},
  journal={Management Science},
  year={2025},
  publisher={INFORMS}
}

@article{buckels2014trolls,
  title={Trolls just want to have fun},
  author={Buckels, Erin E and Trapnell, Paul D and Paulhus, Delroy L},
  journal={Personality and Individual Differences},
  volume={67},
  pages={97--102},
  year={2014},
  publisher={Elsevier}
}

@article{toubia2013intrinsic,
  title={Intrinsic vs. image-related utility in social media: Why do people contribute content to twitter?},
  author={Toubia, Olivier and Stephen, Andrew T},
  journal={Marketing Science},
  volume={32},
  number={3},
  pages={368--392},
  year={2013},
  publisher={INFORMS}
}

@article{liu2022implications,
  title={Implications of revenue models and technology for content moderation strategies},
  author={Liu, Yi and Yildirim, Pinar and Zhang, Z John},
  journal={Marketing Science},
  volume={41},
  number={4},
  pages={831--847},
  year={2022},
  publisher={INFORMS}
}

@article{berger2012makes,
  title={What makes online content viral?},
  author={Berger, Jonah and Milkman, Katherine L},
  journal={Journal of Marketing Research},
  volume={49},
  number={2},
  pages={192--205},
  year={2012},
  publisher={SAGE Publications Sage CA: Los Angeles, CA}
}

@article{craker2016dark,
  title={The dark side of Facebook{\textregistered}: The Dark Tetrad, negative social potency, and trolling behaviours},
  author={Craker, Naomi and March, Evita},
  journal={Personality and Individual Differences},
  volume={102},
  pages={79--84},
  year={2016},
  publisher={Elsevier}
}

@article{jhaver2019did,
  title={" Did you suspect the post would be removed?" Understanding user reactions to content removals on Reddit},
  author={Jhaver, Shagun and Appling, Darren Scott and Gilbert, Eric and Bruckman, Amy},
  journal={Proceedings of the ACM on Human-computer Interaction},
  volume={3},
  number={CSCW},
  pages={1--33},
  year={2019},
  publisher={ACM New York, NY, USA}
}

@article{huang2024politically,
  title={Politically biased moderation drives echo chamber formation: An analysis of user-driven content removals on Reddit},
  author={Huang, Justin T and Choi, Jangwon and Wan, Yuqin},
  journal={Available at SSRN},
  year={2024}
}

@article{wuestenenk2025influence,
  title={The influence of group membership on online expressions and polarization on a discussion platform: An experimental study},
  author={Wuestenenk, Nick and van Tubergen, Frank and Stark, Tobias H},
  journal={New Media \& Society},
  volume={27},
  number={1},
  pages={225--245},
  year={2025},
  publisher={SAGE Publications Sage UK: London, England}
}

@article{berman2020curation,
  title={Curation algorithms and filter bubbles in social networks},
  author={Berman, Ron and Katona, Zsolt},
  journal={Marketing Science},
  volume={39},
  number={2},
  pages={296--316},
  year={2020},
  publisher={INFORMS}
}

@article{cinelli2021echo,
  title={The echo chamber effect on social media},
  author={Cinelli, Matteo and De Francisci Morales, Gianmarco and Galeazzi, Alessandro and Quattrociocchi, Walter and Starnini, Michele},
  journal={Proceedings of the National Academy of Sciences},
  volume={118},
  number={9},
  pages={e2023301118},
  year={2021},
  publisher={National Academy of Sciences}
}

@inproceedings{dave2020social,
  title={Social media giants warn of AI content moderation errors, as employees sent home},
  author={Dave, P},
  booktitle={World Economic Forum. https://www. weforum. org/agenda/2020/03/social-media-giants-a i-moderation-errors-coronavirus/. Accessed December},
  volume={27},
  pages={2020},
  year={2020}
}

@article{levy2021social,
  title={Social media, news consumption, and polarization: Evidence from a field experiment},
  author={Levy, Ro’ee},
  journal={American Economic Review},
  volume={111},
  number={3},
  pages={831--870},
  year={2021},
  publisher={American Economic Association 2014 Broadway, Suite 305, Nashville, TN 37203}
}

@article{muller2021fanning, 
  title={Fanning the flames of hate: Social media and hate crime},
  author={M{\"u}ller, Karsten and Schwarz, Carlo},
  journal={Journal of the European Economic Association},
  volume={19},
  number={4},
  pages={2131--2167},
  year={2021},
  publisher={Oxford University Press}
}

@article{muller2023hashtag,
  title={From hashtag to hate crime: Twitter and antiminority sentiment},
  author={M{\"u}ller, Karsten and Schwarz, Carlo},
  journal={American Economic Journal: Applied Economics},
  volume={15},
  number={3},
  pages={270--312},
  year={2023},
  publisher ={American Economic Association 2014 Broadway, Suite 305, Nashville, TN 37203-2425}
}

@techreport{andres2021combating,
  title={Combating online hate speech: The impact of legislation on Twitter},
  author={Andres, Raphaela and Slivko, Olga},
  year={2021},
  institution={ZEW Discussion Papers}
}

@article{carlson2020report,
  title={Report and repeat: Investigating Facebook’s hate speech removal process},
  author={Carlson, Caitlin Ring and Rousselle, Hayley},
  journal={First Monday},
  year={2020}
}

@article{muller2023effects,
  title={The Effects of Online Content Moderation: Evidence from President Trump's Account Deletion},
  author={M{\"u}ller, Karsten and Schwarz, Carlo},
  journal={Available at SSRN 4296306},
  year={2023}
}

@inproceedings{thomas2022s,
  title={“It’s common and a part of being a content creator”: Understanding How Creators Experience and Cope with Hate and Harassment Online},
  author={Thomas, Kurt and Kelley, Patrick Gage and Consolvo, Sunny and Samermit, Patrawat and Bursztein, Elie},
  booktitle={Proceedings of the 2022 CHI Conference on Human Factors in Computing Systems},
  pages={1--15},
  year={2022}
}

@online{RedditDownvote2025,
  title        = {Reddit Downvote Impact: User Engagement \& Community Dynamics},
  author       = {{Our Mental Health}},
  year         = {2025},
  url          = {https://www.ourmental.health/screen-time-sanity/reddits-downvote-dilemma-how-negative-feedback-shapes-community-engagement},
  note         = {Accessed: 2025-03-16},
}

@article{iyengar2019origins,
  title={The origins and consequences of affective polarization in the United States},
  author={Iyengar, Shanto and Lelkes, Yphtach and Levendusky, Matthew and Malhotra, Neil and Westwood, Sean J},
  journal={Annual Review of Political Science},
  volume={22},
  number={1},
  pages={129--146},
  year={2019},
  publisher={Annual Reviews}
}

@article{kozyreva2023resolving,
  title={Resolving content moderation dilemmas between free speech and harmful misinformation},
  author={Kozyreva, Anastasia and Herzog, Stefan M and Lewandowsky, Stephan and Hertwig, Ralph and Lorenz-Spreen, Philipp and Leiser, Mark and Reifler, Jason},
  journal={Proceedings of the National Academy of Sciences},
  volume={120},
  number={7},
  pages={e2210666120},
  year={2023},
  publisher={National Academy of Sciences}
}

@article{moorthy1985using,
  title={Using game theory to model competition},
  author={Moorthy, K Sridhar},
  journal={Journal of Marketing Research},
  volume={22},
  number={3},
  pages={262--282},
  year={1985},
  publisher={SAGE Publications Sage CA: Los Angeles, CA}
}

@article{grossman1980impossibility,
  title={On the impossibility of informationally efficient markets},
  author={Grossman, Sanford J and Stiglitz, Joseph E},
  journal={The American Economic Review},
  volume={70},
  number={3},
  pages={393--408},
  year={1980},
  publisher={JSTOR}
}

@article{cima2024investigating,
  title={Investigating the heterogeneous effects of a massive content moderation intervention via Difference-in-Differences},
  author={Cima, Lorenzo and Tessa, Benedetta and Cresci, Stefano and Trujillo, Amaury and Avvenuti, Marco},
  journal={arXiv preprint arXiv:2411.04037},
  year={2024}
}

@inproceedings{hickey2023auditing,
  title={Auditing Elon Musk’s impact on hate speech and bots},
  author={Hickey, Daniel and Schmitz, Matheus and Fessler, Daniel and Smaldino, Paul E and Muric, Goran and Burghardt, Keith},
  booktitle={Proceedings of the International AAAI Conference on Web and Social Media},
  volume={17},
  pages={1133--1137},
  year={2023}
}

@article{druckman2019we,
  title={What do we measure when we measure affective polarization?},
  author={Druckman, James N and Levendusky, Matthew S},
  journal={Public Opinion Quarterly},
  volume={83},
  number={1},
  pages={114--122},
  year={2019},
  publisher={Oxford University Press UK}
}

@article{lelkes2017hostile,
  title={The hostile audience: The effect of access to broadband internet on partisan affect},
  author={Lelkes, Yphtach and Sood, Gaurav and Iyengar, Shanto},
  journal={American Journal of Political Science},
  volume={61},
  number={1},
  pages={5--20},
  year={2017},
  publisher={Wiley Online Library}
}

@article{homroy2023political,
  title={Political Polarization and Corporate Political Advocacy},
  author={Homroy, Swarnodeep and Gangopadhyay, Shubhashis},
  journal={Available at SSRN 4742753},
  year={2023}
}

@article{haimson2021disproportionate,
  author    = {Haimson, Oliver L. and Semrau, Mallory and Matias, Nathan and Vitak, Jessica},
  title     = {Disproportionate removals and differing content-moderation experiences for conservative, transgender, and Black social media users},
  journal   = {Proceedings of the ACM on Human–Computer Interaction (CSCW)},
  year      = {2021},
  volume    = {5},
  number    = {CSCW2},
  pages     = {Article 466},
  doi       = {10.1145/3479572}
}

@online{meta2021,
  author       = {Lada, Akos and Wang, Meihong and Yan, Tak},
  title        = {How Does News Feed Predict What You Want to See?},
  year         = {2021},
  month        = jan,
  url          = {https://about.fb.com/news/2021/01/how-does-news-feed-predict-what-you-want-to-see/},
  note         = {Meta Newsroom blog, accessed 30 June 2025}
}

@online{budlight2023,
  author       = {Schad, Carolyn},
  title        = {How the Bud Light boycott started—and why it's still going},
  year         = {2023},
  url          = {https://www.nbcnews.com/news/us-news/bud-light-boycott-explained-rcna90271},
  note         = {NBC News, June 29, 2023},
  urldate      = {2025-06-19}
}

@article{godes2019politics,
  title={Politics, Persuasion and Choice},
  author={Godes, David and Mayzlin, Dina and Camara, Odilon and Chung, Doug and Hydock, Chris and Kotchmar, Richard and Lim, Claire and Moshary, Sarah and Paharia, Neeru and Wernerfelt, Nils and others},
  journal={Available at SSRN 3479876},
  year={2019}
}

@article{thomas2023disrupting,
  author = {Thomas, Daniel Robert and Laila A. Wahedi},
  title = {Disrupting Hate: The Effect of Deplatforming Hate Organizations on Their Online Audience},
  year = {2023},
  journal = {Proceedings of the National Academy of Sciences},
  volume = {120},
  number = {24},
  pages = {e2214080120},
}

@article{noelle1974spiral,
  title={The spiral of silence a theory of public opinion},
  author={Noelle-Neumann, Elisabeth},
  journal={Journal of Communication},
  volume={24},
  number={2},
  pages={43--51},
  year={1974},
  publisher={Oxford University Press}
}

@article{coase1960problem,
  title={The Problem of Social Cost},
  author={Coase, Ronald H.},
  journal={Journal of Law and Economics},
  volume={3},
  number={1},
  pages={1--44},
  year={1960},
  publisher={University of Chicago Press}
}

@misc{birdwatch2021launch,
  author       = {Twitter},
  title        = {Introducing Birdwatch, a community-based approach to misinformation},
  year         = {2021},
  url          = {https://blog.twitter.com/en_us/topics/product/2021/introducing-birdwatch-a-community-based-approach-to-misinformation},
  note         = {Accessed June 30, 2025}
}

@misc{kim2024youtube,
  title        = {The Impact of YouTube’s Hiding Dislike Count on Viewer and Creator Engagement},
  author       = {Kim, Hyejun and Lu, Danqi and Ma, Xinyue and Tafti, Arvind},
  year         = {2024},
  note         = {SSRN Working Paper},
  url          = {https://papers.ssrn.com/sol3/papers.cfm?abstract_id=4967594}
}

@article{lerman2024affective,
  title={Affective polarization and dynamics of information spread in online networks},
  author={Lerman, Kristina and Feldman, Dan and He, Zihao and Rao, Ashwin},
  journal={npj Complexity},
  volume={1},
  number={1},
  pages={8},
  year={2024},
  publisher={Nature Publishing Group UK London}
}

@inproceedings{rajadesingan2021political,
  title={Political discussion is abundant in non-political subreddits (and less toxic)},
  author={Rajadesingan, Ashwin and Budak, Ceren and Resnick, Paul},
  booktitle={Proceedings of the International AAAI Conference on Web and Social Media},
  volume={15},
  pages={525--536},
  year={2021}
}

@article{yu2024partisanship,
  title={Partisanship on social media: In-party love among American politicians, greater engagement with out-party hate among ordinary users},
  author={Yu, Xudong and Wojcieszak, Magdalena and Casas, Andreu},
  journal={Political Behavior},
  volume={46},
  number={2},
  pages={799--824},
  year={2024},
  publisher={Springer}
}

@article{acemoglu2007competition,
  title={Competition and efficiency in congested markets},
  author={Acemoglu, Daron and Ozdaglar, Asuman},
  journal={Mathematics of operations research},
  volume={32},
  number={1},
  pages={1--31},
  year={2007},
  publisher={INFORMS}
}

@article{johari2010investment,
  title={Investment and market structure in industries with congestion},
  author={Johari, Ramesh and Weintraub, Gabriel Y and Van Roy, Benjamin},
  journal={Operations Research},
  volume={58},
  number={5},
  pages={1303--1317},
  year={2010},
  publisher={INFORMS}
}

@article{rieder2021fabrics,
  title={The fabrics of machine moderation: Studying the technical, normative, and organizational structure of Perspective API},
  author={Rieder, Bernhard and Skop, Yarden},
  journal={Big Data \& Society},
  volume={8},
  number={2},
  pages={20539517211046181},
  year={2021},
  publisher={SAGE Publications Sage UK: London, England}
}

@report{pew2024partisanship,
  title        = {The Partisanship and Ideology of American Voters},
  author       = {{Pew Research Center}},
  year         = {2024},
  month        = {April},
  url          = {https://www.pewresearch.org/politics/2024/04/09/the-partisanship-and-ideology-of-american-voters/},
  institution  = {Pew Research Center},
  note         = {Accessed October 29, 2025},
}

@article{gonzalez2023asymmetric,
  title={Asymmetric ideological segregation in exposure to political news on Facebook},
  author={Gonz{\'a}lez-Bail{\'o}n, Sandra and Lazer, David and Barber{\'a}, Pablo and Zhang, Meiqing and Allcott, Hunt and Brown, Taylor and Crespo-Tenorio, Adriana and Freelon, Deen and Gentzkow, Matthew and Guess, Andrew M and others},
  journal={Science},
  volume={381},
  number={6656},
  pages={392--398},
  year={2023},
  publisher={American Association for the Advancement of Science}
}

@article{eg2023scoping,
  title={A scoping review of personalized user experiences on social media: The interplay between algorithms and human factors},
  author={Eg, Ragnhild and T{\o}nnesen, {\"O}zlem Demirkol and Tennfjord, Merete Kolberg},
  journal={Computers in Human Behavior Reports},
  volume={9},
  pages={100253},
  year={2023},
  publisher={Elsevier}
}

@inproceedings{cima2024great,
  title={The great ban: Efficacy and unintended consequences of a massive deplatforming operation on reddit},
  author={Cima, Lorenzo and Trujillo, Amaury and Avvenuti, Marco and Cresci, Stefano},
  booktitle={Companion Publication of the 16th ACM Web Science Conference},
  pages={85--93},
  year={2024}
}

@article{iyer2016competing,
  title={Competing for attention in social communication markets},
  author={Iyer, Ganesh and Katona, Zsolt},
  journal={Management Science},
  volume={62},
  number={8},
  pages={2304--2320},
  year={2016},
  publisher={INFORMS}
}

@misc{Pew2021,
  author       = {{Pew Research Center}},
  year         = {2021},
  title        = {The State of Online Harassment},
  howpublished = {\emph{Pew Research Center}, January 13, 2021},
  url          = {https://www.pewresearch.org/internet/2021/01/13/the-state-of-online-harassment/},
  note         = {Accessed: June 30, 2025}
}


\end{document}